\documentclass[11pt]{article}
\usepackage{amsfonts}
\usepackage[margin=2cm]{geometry}
\usepackage{amsmath,amssymb,amsthm,array,booktabs,cite,dsfont,graphicx,mathtools,multirow,placeins,rotating,stmaryrd,tensor,verbatim,wrapfig}
\usepackage[bookmarksnumbered,linktocpage,pdfstartview=FitH]{hyperref}
\usepackage{hypcap}
\usepackage[margin=2cm]{geometry}
\usepackage{caption}
\usepackage{color}
\usepackage{setspace}
\usepackage{slashed}
\usepackage{changepage}

\newcolumntype{M}[1]{>{$}{#1}<{$}}

\newcommand{\sst}[1]{{\scriptscriptstyle #1}}

\def\0{{\sst{(0)}}}
\def\1{{\sst{(1)}}}
\def\2{{\sst{(2)}}}
\def\3{{\sst{(3)}}}
\def\4{{\sst{(4)}}}
\def\5{{\sst{(5)}}}
\def\6{{\sst{(6)}}}
\def\7{{\sst{(7)}}}

\newcommand{\be}{\begin{equation}}
\newcommand{\ee}{\end{equation}}
\def\ba{\begin{array}}
\def\ea{\end{array}}

\newcommand{\bea}{\begin{eqnarray}}
\newcommand{\eea}{\end{eqnarray}}

\DeclareMathOperator{\Tr}{Tr} 

\DeclareMathOperator{\Pf}{Pf}

\DeclareMathOperator{\SO}{SO}
\DeclareMathOperator{\Orth}{O}

\DeclareMathOperator{\Sp}{Sp}

\DeclareMathOperator{\Un}{U}

\DeclareMathOperator{\Cliff}{Cliff}
\DeclareMathOperator{\Pin}{Pin}
\DeclareMathOperator{\Spin}{Spin}

\newcommand{\blf}[2]{\langle#1 , #2\rangle}
\newcommand{\bn}{\mathbf{n}}

\newcommand{\alg}{\mathds{A}}

\newcommand{\F}{\mathds{F}}
\newcommand{\R}{\mathds{R}}
\newcommand{\C}{\mathds{C}}
\newcommand{\Q}{\mathds{H}}
\newcommand{\Oct}{\mathds{O}}
\newcommand{\Z}{\mathds{Z}}

\begin{document}

\begin{titlepage}
\begin{center}
\hfill DIAS-STP-16-10\\
\hfill IMPERIAL-TP-2016-MJD-04\\
\vskip 2cm

{\huge \bf Majorana Fermions in Particle Physics, Solid State and Quantum Information}

\vskip 1.5cm

{\bf L.~Borsten${}^{1}$ and M.~J.~Duff${}^{2,3}$}

\vskip 20pt
 {\it ${}^1$School of Theoretical Physics, Dublin Institute for Advanced Studies,\\
10 Burlington Road, Dublin 4, Ireland}\\\vskip 5pt

{\it ${}^2$Theoretical Physics, Blackett Laboratory, Imperial College London,\\
 London SW7 2AZ, United Kingdom}\\\vskip 5pt

{\it $^3$Mathematical Institute, University of Oxford,\\
Andrew Wiles Building,
Woodstock Road, Radcliffe Observatory Quarter,\\ Oxford, OX2 6GG, United Kingdom}\\\vskip 5pt

\texttt{leron@stp.dias.ie}\\
\texttt{m.duff@imperial.ac.uk}\\

\end{center}
\vskip 2.2cm

\begin{center} {\bf ABSTRACT}\\[3ex]\end{center}
This review is based on  lectures given by M.~J.~Duff summarising the  far reaching contributions of Ettore Majorana to fundamental physics, with special focus on Majorana fermions in all their guises. The theoretical discovery of the eponymous fermion in 1937  has since had profound implications for   particle physics, solid state and quantum computation. The breadth of these disciplines is testimony to Majorana's genius, which continues to permeate physics today. These lectures offer a whistle-stop tour through some limited subset of the key ideas. In addition to touching on these various applications, we will draw out some fascinating relations connecting the normed division algebras $\R, \C, \Q, \Oct$ to spinors, trialities,   $K$-theory and the classification of stable topological states of symmetry-protected gapped free-fermion systems.
\vspace{0.25in}
\begin{center}\emph{Contribution to the Proceedings of the 53$^\text{rd}$ International School of Subnuclear Physics\\
Erice,  24 June - 3 July 2015 \\ 
Celebration of the Triumph of Ettore Majorana}
\end{center}
\vfill

\end{titlepage}

\newpage \setcounter{page}{1} \numberwithin{equation}{section} \tableofcontents 

\newpage
\section{Introduction: Majorana's vision}\label{intro}
In April 1937,  but eight years into a prolific career tragically cut short just  eleven months later, Ettore Majorana discovered that it is  theoretically consistent for a fermion to be its own anti-particle \cite{Majorana:1937vz}.  Majorana fermions, as they have come to be known in a befitting tribute to their originator,  were at first slow to find their raison d'\^etre, but have since had a profound impact on an impressive array of seemingly disparate fields including  particle physics, condensed matter    and  quantum  computing. They are also crucial to the notion of supersymmetry and hence  related approaches to unification and quantum gravity such as M-theory. This particular story is described in the sister contribution dedicated to Majorana \cite{Ferrara:2015bqa}  appearing in these same proceedings. Here instead we attempt  to review Majorana fermions in particle physics, solid state and quantum information, guided chiefly by the excellent accounts given in \cite{adams1996lectures, harvey1990spinors, Baez:2001dm, Nayak:2008zza, alicea2012new, Beenakker:2013, Elliott:2014iha, Akhmedov:2014kxa, Wilczek:2014lwa,Beenakker:2015, chiu2015classification}. Although all are tied tightly to Majorana's pioneering insights some examples rather stretch the original incarnation of 1937 and others even the very notion of a fermion. It would therefore serve us well to  begin, more modestly, as  Majorana did.  

When Dirac first formulated his equation for a relativistic electron\footnote{Here we adopt west coast conventions.},
\be\label{dirac}
(i\gamma^\mu\partial_\mu - m )\psi=0, \qquad \{\gamma^\mu, \gamma^\nu\}= 2\eta^{\mu\nu},
\ee
he hit  upon a realisation of the four matrices  $\gamma^\mu$ containing both real and imaginary elements, the implication being that $\psi$ itself must be complex valued. What a triumph; this single observation anticipated the existence of antimatter. Electrons and positrons are simultaneously  accommodated by $\psi$ and $\psi^*$ distinct.  This success, however, does not imply the mathematical or theoretical necessity of a complex $\psi$. Motivated in part  by mathematical elegance Majorana  questioned this apparent implication and in doing so discovered an alternative realisation in which the four $\gamma^\mu$  are all pure imaginary and hence $\psi$ may be consistently taken as real valued. Pure imaginary $\gamma$ and real $\psi$ are referred to as a ``really real'' Majorana representation and constitute  a special, but entirely equivalent, case of the more general reality condition used to define Majorana fermions,
\be
\overline{\psi}^D=\overline{\psi}^M,
\ee
where $\overline{\psi}^D$ and $\overline{\psi}^M$ denote the Dirac and Majorana conjugates, respectively.

Projecting \eqref{dirac} with the usual chiral operator,
$
P_R=\frac{1}{2}\left(1+\gamma_5\right),
$
in the really real Majorana representation we obtain 
\be
i\gamma^\mu\partial_\mu \psi_{L} - m \psi_{L}^{*}=0,
\ee
where we have used $P_R\psi=P_{L}^{*}\psi=(P_{L}\psi)^{*}$. Hence,   the global phase symmetry $\psi\rightarrow e^{i\theta}\psi$ enjoyed by the Dirac equation is broken by the mass term. Consequently, massive Majorana fermions cannot be coupled to a $\Un(1)$ gauge potential and are  uncharged. Note, for commuting variables the kinetic and mass terms of the  action, 
\be\label{ml}
S[\psi]=-\frac{1}{2}\int \bar{\psi}(i\gamma^\mu\partial_\mu - m )\psi,
\ee
 are symmetric and antisymmetric, respectively.  Hence, the kinetic term is a total derivative while the mass term vanishes  identically, rendering the action trivial.  This led Majorana to invoke Grassmann valued fields. That these are now a pedestrian  part of our mathematical description  of fermions again speaks to Majorana's vision.

We conclude: the tenets of relativity and quantum theory allow, in principle, for neutral fermions that are their own anti-particles. This of course leaves their relevance to fundamental physics unaddressed. Majorana himself speculated that such fermions could represent neutrinos on the basis that they do not carry any electromagnetic charge, but at the time of writing neutrinos were largely hypothetical and little of their would-be properties was known. The picture of neutrinos that evolved over the subsequent years appeared to rule out Majorana's tentative hypothesis. In the standard model of particle physics  they are described by  massless Weyl spinors with distinct anti-particles.  However, this framework has since been eroded by neutrino oscillations \cite{Fukuda:1998mi}, which  imply, at least without some new and rather radical ingredients, that  neutrinos carry a small mass.  The off-diagonal mass terms required for neutrino oscillations (or any mass terms for that matter) are incompatible with the hypothesis that neutrinos are Weyl fermions. 

Of course, this does not necessarily mean that neutrinos are Majorana as opposed to Dirac, but there are processes which could in principle distinguish the two possibilities, such as double beta decay. See  \cite{Avignone:2007fu} and the references therein. Single beta decay is energetically forbidden in most nuclei with even atomic number $Z$ and even neutron number $N$. However, a large class of even $Z$, even $N$ nuclei do allow the second order process of double beta decay, accompanied with the emission of two anti-neutrinos  conserving lepton number. This process has been observed in a number of isotopes and is consistent with the standard model. If, however, neutrinos are Majorana then there is another possible double beta decay process that does not produce any neutrinos and hence changes the lepton number by two. If neutrinos are their own anti-particles then beta decay is possible via either  the emission or absorption equivilantly. Hence, double beta decay can be mediated by two neutrons exchanging a single neutrino, a process forbidden within standard model. Although there exist a number of alternative mechanisms, the observation of neutrino-less double beta decay would unambiguously imply neutrinos are Majorana \cite{Schechter:1981bd}. The decay rate of  Majorana neutrino mediated neutrino-less double beta decay is quadratic in the effective Majorana neutrino mass, consistent with the observation that  in the massless limit Dirac and Majorana neutrinos are in general physically indistinguishable.  Although neutrino oscillation measurements cannot separate Dirac and Majorana neutrinos, they do provide some   constraints on the effective Majorana neutrino mass. If the heavier masses observed in early measurements of atmospheric neutrino oscillations provide the  dominant contribution, then the effective Majorana neutrino mass would be expected to fall somewhere between 15 and 50 MeV. The current generation of double beta decay experiments should probe the Majorana hypothesis down to an effective Majorana mass of around 100 MeV while the next generation is anticipated to cover the full range of 15-50 MeV in the coming decade.  A nice summary of these experimental efforts is given in \cite{Elliott:2014iha}.

In addition to having some  promising experimental prospects, the Majorana neutrino hypothesis is also phenomenologically attractive. In particular, the seesaw mechanism   \cite{Minkowski:1977sc,Mohapatra:1979ia, Yanagida:1979as, Glashow:1979nm, GellMann:1980vs} simultaneously accounts for light left-handed and very heavy right-handed neutrinos, as required by their non-observation, while providing a consistent mechanism for leptogenesis. Could it be that Majorana left us with an important piece of the puzzle explaining the asymmetry between matter and anti-matter and hence our origins? Evidently, there are  some provocative reasons to believe that Majorana's original plan for his fermion may yet be vindicated. 

Regardless, it is often the case that particularly profound insights  inspire developments that go well beyond the intended domain of application. Majorana fermions are no exception and in the following we will explore some these unanticipated pay-offs.   This is a nice example of how curiosity driven pure science often pays its way many times over; ask good scientific questions and good answers, more broadly understood, will follow.  We will begin in \autoref{cliff} by reviewing spinors from a slightly more mathematical perspective, preparing the ground for some fascinating  connections described in \autoref{mzm}.  We will then give two examples of Majorana's idea appearing in solid state physics. First in \autoref{scmf}  we discuss emergent Majorana fermions as quasiparticle states in superconductors. Second,  we explore the idea of Majorana zero modes (Mzm) in \autoref{mzm}. Although Mzm are, in the spirit of Majorana, their own antiparticles, they obey anyonic statistics and we therefore avoid using the term fermion.  This  feature  makes Mzm an exciting prospect for fault-tolerant  quantum computing as we discuss in \autoref{qc}. We will then discuss the definition and classification of topological phases for gapped free-fermion systems more generally in \autoref{topclass}.  This leads us to some fascinating relations connecting spinors,  normed division algebras,  $K$-theory and the classification  of stable topological states of symmetry-protected gapped free-fermion systems \cite{PhysRevB.78.195125, Kitaev:2009mg, Ryu:2010, Chiu:2011, Freed:2012uu, de2014classification,de2015classification, Kennedy:2015,Kennedy:2014cia, de2015chiral, Thiang:2014fxa}.  In this manner we go full circle, starting from Majorana's work on spacetime spinors, then onto Mzm in condensed matter, their place in the classification of topological phases, and finally back to the theory of spinors via $K$-theory and division algebras. Octonions, the largest of the four normed division algebras,  have appeared in fundamental physics  in a variety of guises, see for example \cite{Gunaydin:1975mp, Gursey:1978et, Julia:1980gr, Kugo:1982bn, Gunaydin:1983rk,  Sudbery:1984, Baez:2001dm, Borsten:2008wd,  Baez:2009xt, Borsten:2013bp, Anastasiou:2013cya, Anastasiou:2014zfa, Borsten:2015pla} and the references therein, primarily through their intriguing connections to spacetime geometry, supersymmetry and exceptional Lie groups. Here we see another, rather subtle, occurrence in the context of condensed matter systems. It is not yet clear, however, whether or not the octonions enter the physics in any meaningful way, let alone if this perspective sheds new light on the problem.   See \autoref{div} for a lightening tour of the division algebras. 

In light of the broad and interdisciplinary scope of such a review we have sought to be as elementary as possible. Our hope is that someone unfamiliar with the various topics covered will  at least be able to take home some limited intuition     without too much further reading. Given our space constraints we are, however, doomed to failure in this regard, for which we apologise in advance. For the same reason we regretfully will have to omit numerous important topics, but hopefully the references provided will help in this regard.

\section{Spinors, trialities and algebras}\label{cliff}

We shall begin by briefly recalling the theory of spinors from the perspective of Clifford algebras as laid out in \cite{Atiyah:1964, harvey1990spinors}. We shall then describe how these structures, together with the notion of  triality \cite{adams1996lectures},  lead us to the four normed division algebras, $\alg=\R, \C, \Q, \Oct$ as described in \cite{adams1996lectures, Baez:2001dm}.  For the bare essentials on the division algebras see \autoref{div}. With this relation in hand we shall outline the connections between normed division algebras, Bott periodicity and real $K$-theory  with a view to their application to topological phases. The    interconnected relations shared by Clifford algebras, Bott periodicity and $K$-theory constitute an elegant and well-known story described in a number  of excellent treatises, such as \cite{Husemoller:1993,karoubi2008k}.  The special role played by the octonions  was  emphasised by, in particular,  Adams and Baez in \cite{adams1996lectures, Baez:2001dm}, which we follow closely here.

For an $d$-dimensional  pseudo-Euclidean vector space $V$ with inner-product $\langle, \rangle$ of signature $s+t=d$ we define  the Clifford algebra on $V$, denoted $\Cliff(V)$, as the   associative algebra freely generated by $V$ modulo the Clifford relations,
\be
xy+yx=-2\langle x, y\rangle
\ee
for all $x,y\in V$. Note, the opposite sign convention is taken in \cite{Atiyah:1964}. As a vector space (but not as an algebra) 
\be
\Cliff(V) \cong \bigoplus_{k=0}^{d}\wedge^k V,
\ee 
and so $\dim \Cliff(V)=2^d$.
If $V=\R^{s,t}$ with the canonical inner-product we will write $\Cliff(s,t)$. 

The representation theory  is underpinned by the fundamental lemma of Clifford algebras: Suppose $\mathcal{A}$ is a unital associative algebra. Then any linear map $\phi:V\rightarrow\mathcal{A}$ such that,
\be
\phi(x)\phi(x)=-\langle x, x\rangle\mathds{1}, \quad \forall x\in V,
\ee
admits a unique extension to an algebra homomorphism, $\hat{\phi}:\Cliff(V) \rightarrow\mathcal{A}$.

For a unital algebra $\mathcal{A}$ let $\mathcal{A}[n]$ denote  the set of $n\times n$ matrices with entries in $\mathcal{A}$. All Clifford algebras $\Cliff(s,t)$ are isomorphic to  the (direct sum of) matrix algebras  $\alg[n]$ for $\alg=\R,\C, \Q$, as given in \autoref{tab:AD}. Here we already see the Bott periodic (mod 8 repetition) features discovered by Cartan \cite{Cartan:1908},
\be
\Cliff(s+8, t)\cong \Cliff(s, t+8)\cong\Cliff(s, t)\otimes \R[16], \quad\text{where}\quad \Cliff(8,0)\cong\Cliff(0,8)\cong \R[16].
\ee
 Note, starting from $s-t=5$, say, we have the  useful mnemonic:
\be\label{mn}
\C\qquad   \R\quad    \R\R\quad    \R\qquad    \C \qquad    \Q\quad    \Q\Q\quad    \Q\qquad    \C\qquad    \R\quad    \R\R\quad    \R\qquad    \C \qquad    \Q\quad    \Q\Q\quad    \Q\ldots
\ee
If we complexify this pattern simplifies. Since all complex inner product spaces with the same dimension $d=s+t$ are isometric and the isometry lifts to a unique Clifford algebra automorphism there is a unique complex Clifford algebra for a given dimension, which is given by
\be
\Cliff_\C(d)=\Cliff(s,t)\otimes_\R \C.
\ee
The order eight pattern of the $\Cliff(s,t)$ matrix representations given in \autoref{tab:AD} is reduced to an order two ``Bott periodic'' sequence:
\be\label{CCliff}
\begin{array}{ccccccccc}
d\mod 2 && \Cliff_\C(d) &&\Pin_\C(d) ~\text{irreps} \\[5pt]
0 && \C[n] &&\C^n \\[5pt]
1 && \C[n]\oplus \C[n]  &&\C_+[n] \quad \C_-[n] \\
\end{array}
\ee
where $n$ is fixed by the dimension of $\Cliff_\C(d)$, $n=2^{d/2}$ for $d=0\mod 2$ and $n=2^{(d-1)/2}$ for $d=1\mod 2$.

The  isomorphism with matrix algebras implies that the non-trivial  $\Cliff(s,t)$-modules are  isomorphic to either      $\alg^n$ or  $\alg^n\oplus\alg^n$, as given in \autoref{tab:AD}, and $\C^n$ or  $\C^n\oplus\C^n$ in the complexified case. These are referred to as pinor representations for reasons that will become clear shortly. 
\newcolumntype{M}{>{$}c<{$}}
\begin{table}[h]
\centering
\begin{tabular}{M|M|M|M }
 \hline
 \hline
&&& \\
s-t \,\text{mod} \, 8&\Cliff(s,t)\cong\Cliff_0(s+1, t) & \Pin(s,t) ~\text{irreps}~ P&\Spin(s,t) ~\text{irreps} ~ S \\[5pt]
\hline
&&& \\
0& \R[n] 			& \R^n	&\R^{n}_+\quad \R^{n}_{-} \\[5pt]
1& \C[n] 			& \C^n 	&\R^n  \\[5pt]
2&\Q[n]			&\Q^n 	&\C^n \\[5pt]
3&\Q[n]\oplus \Q[n]  	&\Q^{n}_{+}\quad \Q^{n}_{-} &\Q^n\\[5pt]
4 &\Q[n]& \Q^{n}  	&\Q^{n}_{+}\quad \Q^{n}_{-}  \\[5pt]
5 & \C[n]			&\C^n 	& \Q^n  \\[5pt]
6 &\R[n]			&\R^n &\C^n  \\[5pt]
7 & \R[n]\oplus \R[n] &\R^{n}_{+} \quad\R^{n}_{-}&  \R^n    \\
&&& \\
 \hline
 \hline
\end{tabular}
\caption{The Clifford algebras and  minimal pin and spin representations. The  parameters $n$ for the pinors are determined by the real dimension of the Clifford algebra $\Cliff(s,t)$ and for the spinors by the the real dimension of the even Clifford algebra $\Cliff_0(s,t)\cong \Cliff(s-1,t)$, where $\dim_\R \Cliff(s,t)$ is given by $n^2\dim \alg=2^{d}$ for $s-t \, \text{mod}\, 8= 0,1,2,4,5,6$ and $2n^2\dim \alg=2^{d}$ for $s-t \, \text{mod}\, 8= 3,7$. }\label{tab:AD}
\end{table}
In particular, we have
\be
\Cliff(0,0)\cong\R[1], \qquad \Cliff(1,0)\cong\C[1], \qquad \Cliff(2,0)\cong\Q[1].
\ee 
and their corresponding pinor representations, $\R, \C, \Q$. Although $\Oct$, being non-associative (see \autoref{div}), is not itself a Clifford algebra, it does form a representation of $\Cliff(\text{Im}\Oct)\cong\Cliff(7, 0)$ \cite{adams1996lectures}. 
More generally, for a normed division algebra $\alg$  consider the linear map, 
\be
\begin{array}{ccccccccc}
\phi : &\text{Im}\alg&\rightarrow&L(\text{Im}\alg),\\
&a&\mapsto& L_a
\end{array}
\ee 
where we have defined the left multiplication operator $L_a(x):=ax, \forall a,x\in \alg$. Since $\text{Im}\alg$ can be regarded as the tangent space to the unit sphere in $\alg$ and $L_a$ is norm-preserving for $||a||=1$, the set of  maps in $L(\text{Im}\alg)$ are skew-symmetric and satisfy,
\be
L_{a}L_{a} = -\langle a, a\rangle.
\ee
By the fundamental lemma of Clifford algebras we therefore have a faithful (as $\text{Im}\alg$ is orthogonal to the span of the identity in $\alg$) representation of $\Cliff(\text{Im}\alg)$ on $\alg$.

Let us now turn to the defintion of the pin and spin groups, $\Pin(s,t)$ and $\Spin(s,t)$, which sit inside $\Cliff(s,t)$. The involution on $V$ given by $\tilde{v}=-v, v\in V$ induces an  automorphism on $\Cliff(V)$. Let the even part of $\Cliff(V)$ be defined by,
\be
\Cliff_0(V) := \{x\in\Cliff(V)\, | \,\tilde{x}=x\}.
\ee
For $V\cong\R^{s,t}$ we then have the canonical Clifford algebra isomorphisms \cite{harvey1990spinors}:
\be
\begin{array}{llllllllllllll}
&\Cliff(s-1,t)&\cong&\Cliff(t-1,s),\\[5pt]
&\Cliff_0(s,t)&\cong&\Cliff(s-1,t),\quad s\geq1\\[5pt]
&\Cliff_0(s,t)&\cong&\Cliff(t-1, s),\quad t\geq1
\end{array}\label{cliff_isos}
\ee
which imply 
\be
\Cliff_0(s,t)\cong\Cliff_0(t,s).
\ee

The multiplicative group $\Cliff^*(s,t)$ in $\Cliff(s,t)$ is given by,
\be
\Cliff^*(s,t):=\{x\in\Cliff(s,t)\,|\, x=v_1\cdots v_r, \quad \langle v_i, v_i\rangle\not=0, v_i\in V \}.
\ee
The  double cover $\Pin(s,t)$ of  $\Orth(s,t)$ is given by the pin subgroup in $\Cliff^*(s,t)$ generated by unit vectors,
\be
\Pin(s,t):=\{x\in\Cliff^*(s,t)\,|\, x=v_1\cdots v_r, \quad |\langle v_i, v_i\rangle|=1, v_i\in V \}.
\ee
Similarly, the spin group $\Spin(s,t)$ double covers  $\SO(s,t)$ and is given by the  subgroup in $\Cliff^*(s,t)$ generated by an even number of unit vectors,
\be
\Spin(s,t):=\Pin(s,t)\cap \Cliff_0(s,t).
\ee
From these definitions it follows  that the irreducible  representations of $\Cliff(s,t)$ and $\Cliff_0(s,t)$ restrict to irreducible representations of $\Pin(s,t)$ and $\Spin(s,t)$, respectively, as given in  \autoref{tab:AD}. Including the conjugate representations  for the complex cases completes the classification of all pinor and spinor representations in all signatures as real, complex or quaternionic vector spaces. 

 The standard nomenclature used in high energy physics typically applies the term ``spinor'', with various qualifiers (Dirac, Majorana, Weyl, symplectic  etc),  to cover all cases. See for example the discussion in \cite{figueroa2006majorana} used here. In particular,  ``Majorana spinor'' as used in \autoref{intro} corresponds to the pinor representation of $\Cliff(1, 3)$, which according to \autoref{tab:AD} is isomorphic to $\R^4$, as we would expect. To reconnect to the more familiar path to Majorana spinors, note that the  complex pinor representation $P_\C\cong\C^4$ of the complexified Clifford algebra $\Cliff_\C(4)\cong\Cliff(1, 3)\otimes_\R\C$ admits an invariant  real structure: a linear map  $\varphi:P_\C\rightarrow P_\C$ such that $\varphi(\alpha \psi)=\alpha^* \varphi(\psi)$ and $\varphi^2=1$.  The  elements in $P_\C$ invariant under $\varphi$ are said to be Majorana.  This is  equivalent to the usual definition that the Dirac  and Majorana conjugates are identical on Majorana spinors.  This follows from the fact that given an invariant Hermitian form $\langle -, - \rangle: P_\C\times P_\C\rightarrow\C$ (which always exists) and an invariant  real structure on $P_\C$ one can construct a spin invariant  non-degenerate complex  symmetric bilinear form $B:P_\C\times P_\C\rightarrow\C$,
 \be
 B(\psi, \chi):=\langle \varphi(\psi), \chi \rangle.
 \ee
On    Majorana spinors  $\varphi(\psi)=\psi$ we therefore have $B(\psi, \chi)=\langle \psi, \chi \rangle$ for all $\chi$. In a basis-dependent form
\be
\psi^TB\chi = \psi^\dagger A\chi, \qquad \forall \chi,
\ee
where $B,A$  are symmetric and hermitian matrices, respectively. Hence
\be
\overline{\psi}^M:=\psi^TB = \psi^\dagger A=:\overline{\psi}^D,
\ee
where the left-hand side defines the Majorana conjugate,  the matrix $B$ is the charge conjugation matrix typically denoted $C$, while the right-hand side defines the Dirac conjugate. Since the  usual irreducible conjugate Weyl spinors, $S$ and $\overline{S}$, of $\Spin(1,3)$ (which are equivalent as  $\R$-representations under the intertwiner given by conjugation on $P_\C$) can be obtained from $P_\C$ by projecting on the positive and negative eigenspaces of a unit volume element, $S\cong \{x \in P_\C \, |\, \lambda x = ix\}$ and $\overline{S}\cong \{x \in P_\C \, |\, \lambda x = -ix\}$, we can regard $P_\C$ as their direct sum. The Majorana condition then relates the two Weyl spinors in the familiar manner.

We have already seen that the four normed division algebras, $\R, \C, \Q$ and $\Oct$, form representations of the Clifford algebras $\Cliff(\text{Im}{\R}), \Cliff(\text{Im}{\C}), \Cliff(\text{Im}{\Q})$ and $\Cliff(\text{Im}{\Oct})$. We would now like to make the connection between $\R, \C, \Q, \Oct$ and spinors even more concrete, using the notion of triality \cite{adams1996lectures, Baez:2001dm}. Here we focus on the Euclidean case $(s,t)=(d,0)$. The group $\Pin(d)$ has, besides the pinor representations $P$, a vector representation $V\cong\R^{d}$. The obvious inclusion $V\hookrightarrow\Cliff(d)$ allows us to restrict the Clifford algebra action on $P$ to an intertwiner of $\Pin(d)$ representations,
\be
\begin{array}{llllllllllllllllllll}
\tilde{m}:& V \times P^{\pm}&\rightarrow &P^{\mp} & d=3,7\,\text{mod} \,8\\[5pt]
\tilde{m}:& V \times P&\rightarrow &P & \text{otherwise} 
\end{array}
\ee
Decomposing the  pinor representations under $\Spin(d)$ we obtain
 \be
\begin{array}{llllllllllllllllllll}
\tilde{m}:& V \times S^{\pm}&\rightarrow& S^{\mp} & d=0,4\,\text{mod} \,8\\[5pt]
\tilde{m}:& V \times S&\rightarrow& S & \text{otherwise} 
\end{array}
\ee 
All such spinor spaces admit an inner product (see for example \cite{harvey1990spinors}) and so can be dualised to give trilinear maps  
 \be\label{ptri}
\begin{array}{llllllllllllllllllll}
\tilde{t}:& V \times S^{\pm}\times S^{\mp}&\rightarrow &\R & d=0,4\,\text{mod} \,8\\[5pt]
\tilde{t}:& V \times S\times S&\rightarrow &\R & \text{otherwise} 
\end{array}
\ee 
 which we could refer to as ``pre-trialities'', for reasons that we will now make clear.  
 
Given three  real vector spaces  a triality is a non-degenerate trilinear map, 
\be
t:V_1\times V_2\times V_3\rightarrow \R.
\ee
We can dualise to produce a bilinear map
\be
m:V_1\times V_2 \rightarrow V_{3}^{*},
\ee
which can be interpreted as left (right) multiplication by elements in $V_1$ ($V_2$). The non-degeneracy of $t$ then implies a set of isomorphisms $V_1\cong V_2\cong V_3\cong V$ such that  $m:V\times V \rightarrow V$ defines a division algebra; $m$ really does stand for multiplication. Conversely, all division algebras define a triality. In particular, a normed division algebra gives a normed triality  satisfying,
\be
|t(x,y,z)|\leq \| x\| \, \|y\| \, \|z\|,
\ee
where  for all $x,y$ there is a $z$ such that the bound is saturated. Conversely, any normed triality gives a normed division algebra. 

We now can appreciate  the key observation: the pre-trilaities  defined on the vector and spinor representations in \eqref{ptri} will yield bona fide normed trialties if and only if $\dim V = \dim S^{(\pm)} = d$. Consulting \autoref{tab:AD}, we see that this happens precisely for $d=1,2,4,8$:
\ \be\label{tri-alg}
\begin{array}{lllllllllllllll}
t_1 :& V_1 \times S_{1}\times S_{1}&\rightarrow&\R&\Leftrightarrow &\R  \\[5pt]
t_2 :& V_2 \times S_{2}\times S_{2}&\rightarrow&\R&\Leftrightarrow &\C  \\[5pt]
t_4 :& V_4 \times S_{4}^{+}\times S_{4}^{-}&\rightarrow&\R&\Leftrightarrow &\Q  \\[5pt]
t_8 :& V_8 \times S_{8}^{+}\times S_{8}^{-}&\rightarrow&\R&\Leftrightarrow &\Oct  \\[5pt]
\end{array}
\ee 
In particular, the octonions provide a representation for the three  8-dimensional representations  of $\Spin(8)$.

We have now seen two ways in which the the normed division algebras are connected to Clifford algebras and their representations. The number eight appears to be omnipresent and one might wonder if there is a deeper connection between the Bott periodic patterns displayed by  Clifford algebras and  the appearance of the normed division algebras, in particular the 8-dimensional octonions. Indeed there is and in more ways than one. Of particular interest to us is the relation mediated by real $K$-theory. 

Before we come to that, let us take a look at  Bott periodicity through the division algebraic lens \cite{Baez:2001dm}. We have  a   relationship to $(\dim \alg)$-dimensional spheres via the division algebraic projective lines treated as smooth manifolds,
\be
\alg\mathds{P}^1 \cong S^{\dim \alg}.
\ee
One has to take extra care for $\Oct\mathds{P}^1$ when defining ``a line through the origin'', see for example the treatment of the Cayley plane given in \cite{harvey1990spinors},  but otherwise it is conceptually entirely equivalent to the familiar Riemann sphere $\C\mathds{P}^1$. Now, consider the maps $f_\alg$, as defined  in  \cite{Baez:2001dm}, sending norm-one elements  $a\in\alg$ to the orthogonal linear operators on $\alg$ given by right-multiplication $R_a(x)=xa$,
\be
\begin{array}{ccccccccc}
f_\alg: &S^{\dim \alg -1}&\rightarrow&\text{O}(\dim \alg),\\
&a&\mapsto& R_a
\end{array}
\ee 
Any real  vector bundle over $S^{\dim \alg}$ with $(\dim \alg)$-dimensional fibres can be form by gluing the trivial bundles on the northern and southern hemispheres along the equator using a map $f: S^{\dim \alg -1}\rightarrow \Orth(\dim\alg)$.  See for example \cite{karoubi2008k}. For $f=f_\alg$ we obtain the canonical line bundles $L_\alg$ \cite{Baez:2001dm}; the fibre at a point in $\alg\mathds{P}^1$ is  the corresponding line through the origin, forming a copy of $\alg$. The homotopy classes of the  maps $f_\alg$, denoted $[f_\alg]$, generate the non-trivial  homotopy groups of the topological group $\Orth\equiv\Orth(\infty)= \text{inj} \lim \Orth(n)$, 
\be\label{pi}
\begin{array}{llllllllll}
[f_\R]& \rightarrow &\pi_0 [\Orth(\infty)]&\cong& \Z_2, \\[5pt]
[f_\C]& \rightarrow &\pi_1 [\Orth(\infty)]&\cong& \Z_2, \\[5pt]
&  &\pi_2[\Orth(\infty)] &\cong& \varnothing, \\[5pt]
[f_\Q]& \rightarrow &\pi_3 [\Orth(\infty)]&\cong& \Z, \\[5pt]
&  &\pi_4 [\Orth(\infty)]&\cong& \varnothing, \\[5pt]
&  &\pi_5 [\Orth(\infty)]&\cong& \varnothing, \\[5pt]
&  &\pi_6[\Orth(\infty)] &\cong& \varnothing, \\[5pt]
[f_\Oct]& \rightarrow &\pi_7[\Orth(\infty)] &\cong& \Z, \\
\end{array}
\ee
as computed by Bott \cite{Bott:1957}, who also established the  periodic behaviour bearing his name,
\be\label{bott}
\pi_i[\Orth(\infty)]\cong\pi_{i+8}[\Orth(\infty)].
\ee
Note, the non-trivial cases occur for $i=\dim\alg -1$ mod 8. Of course, this follows from the fact that $f_\alg$ defines a map from $S^{\dim\alg-1}$ to $\Orth(\dim\alg)\hookrightarrow\Orth(\infty)$ \cite{Baez:2001dm}. 

In 1961 Atiyah and Hirzebruch showed \cite{Atiyah:1961} that Bott periodicity \eqref{bott} is related to  algebraic $K$-theory.  In particular, the sequence of discrete groups \eqref{pi} is  replicated by the reduced real $K$-theory of $n$-spheres for $n=1,2\ldots 8$:
\be\label{rrk}
\widetilde{KO}(S^n)\cong\pi_{n-1}[\Orth(\infty)].
\ee
 Here, the nontrivial cases correspond to $S^{\dim\alg}$ and the normed division algebras are again, from a certain perspective, responsible. To describe this observation, let us briefly recall the basic notions behind $K$-theory. $K$-theory is a deep set of ideas that we cannot possibly do justice to here; hopefully our very limited sketch will be enough to develop some heuristic intuition. The interested reader is encouraged to consult as a starting point the excellent introductions \cite{Atiyah:1967}\footnote{This focusses on complex $K$-theory, but the key concepts largely transfer to the real case.} and \cite{karoubi2008k}.

Real $K$-theory probes the structure of a topological space $X$ by  constructing a ring from the set of  isomorphism classes,  $\text{Vect}(X)$, of real vector bundles over $X$.  Note, $\text{Vect}(X)$ is already a semi-ring (i.e.~a ring with  the axiom of  additive inverses relaxed) with addition and multiplication  given by the direct sum and  tensor product, respectively, of representative bundles. This semi-ring can be completed to produce a ring\footnote{In general, the multiplicative composition  of the semi-ring is not required and we can consider instead an Abelian monoid, an Abelian group with the requirement of the existence of inverses relaxed. Then we have instead the weaker, in that we do not consider any multiplicative structure, group completion with the same conditions as the ring  completion otherwise. The set of isomorphism classes of objects in an additive category $\mathcal{C}$ forms an Abelian  monoid. The group completion in this case is referred to as the Grothendieck group, denoted $K(\mathcal{C})$, and constitutes the fundamental example of this notion. See \cite{karoubi2008k} for a proper account of these constructions.}. A familiar  example of this procedure is given by the construction of the   integers $\Z$ from the  natural numbers $\mathds{N}$. More generally, the ring completion of a semi-ring $R_0$ is a pair $(R, \tau)$, where $R$ is a ring and $\tau$ a semi-ring homomorphism $\tau: R_0\rightarrow R$ such that for any semi-ring  homomorphism $f:R_0\rightarrow S$, $S$ a ring, there exists a unique homomorphism $g:R\rightarrow S$ satisfying $g\circ\tau=f$. One construction of $(R, \tau)$ is given by $R=R_0\times R_0/\sim$, where 
\be
(a,b)\sim(a', b') \quad \Leftrightarrow \quad \exists c\in R_0 \quad \text{s.t.} \quad a+b'+c=a'+b+c,
\ee
and $\tau(a)=[(a,0)]$. The additive identity is $[(0,0)]$ and a quick check confirms that the inverse of $[(a,b)]$ is $[(b,a)]$. Applying this construction to $\text{Vect}(X)$ we obtain a ring denoted $KO(X)$, which is referred to as the real $K$-theory of $X$. For a pointed space $X$,  the homomorphism $rk_0$ sending an isomorphism class in $\text{Vect}(X)$  to the dimension of a representative  fibre at the basepoint of $X$ extends, by construction, to a homomorphism $rk : KO(X)\rightarrow \Z$ such that $rk\circ\tau=rk_0$. The reduced real $K$-theory of $X$ that we alluded to in \eqref{rrk}  is then defined as the kernel of $rk$ and denoted $\widetilde{KO}(X)$. A brief remark on notion: it is often convenient to denote the $K$-theory of complex and real vector bundles by $K_\C$ and $K_{\R}$ (instead of $K$ and $KO$), respectively, reserving the bare $K$ for the  unspecified case.  

An $n$-sphere $S^n$ with the north pole marked constitutes a pointed space and the reduced real $K$-theory for $n$ mod 8 is  given by  
\be\label{pi}
\begin{array}{llllllllll}
[L_\R]& \rightarrow &\widetilde{KO}(S^1)&\cong& \Z_2, \\[5pt]
[L_\C]& \rightarrow &\widetilde{KO}(S^2)&\cong& \Z_2, \\[5pt]
&  &\widetilde{KO}(S^3) &\cong& \varnothing, \\[5pt]
[L_\Q]& \rightarrow &\widetilde{KO}(S^4)&\cong& \Z, \\[5pt]
&  &\widetilde{KO}(S^5)&\cong& \varnothing, \\[5pt]
&  &\widetilde{KO}(S^6)&\cong& \varnothing, \\[5pt]
&  &\widetilde{KO}(S^7)&\cong& \varnothing, \\[5pt]
[L_\Oct]& \rightarrow &\widetilde{KO}(S^8)&\cong& \Z, \\
\end{array}
\ee
where we have indicated that the non-trivial cases $\widetilde{KO}(S^{\dim \alg})$ are generated by the isomorphism class of the canonical $\alg$-line bundle \cite{Baez:2001dm}. Hence $\widetilde{KO}(S^n)\cong\pi_{n-1}[\Orth(\infty)]$. We can moreover construct  a $\Z$-graded ring,
\be
\widetilde{KO}=\bigoplus_n \widetilde{KO}(S^n),
\ee
with multiplication induced by smash products of vector bundles. Multiplication by $[L_\Oct]$ gives an isomorphism $\widetilde{KO}(S^n)\cong \widetilde{KO}(S^{n+8})$  \cite{Baez:2001dm}. In this sense it is the octonions that generate Bott periodicity!

Of course, the shared Bott periodic features of Clifford algebras, spinors, division algebras and real $K$-theory are no coincidence.  In particular, Atiyah, Bott and Shapiro made a detailed instigation \cite{Atiyah:1964} into  Clifford algebras, spinors and real $K$-theory. First of all, let $M_k$ denote the Abelian group freely generated by the irreducible $\Z_2$-graded $\Cliff(k)$-modules. The isomorphism $\hat{\phi}:\Cliff(k+l)\rightarrow\Cliff(k)\otimes\Cliff(l)$ given by the linear extension of 
\be
\phi(e_i)=
  \begin{cases}
    e_i\otimes \mathds{1}       & \quad i\leq k\\
    \mathds{1}\otimes e_i & \quad k< i\\
  \end{cases}
  \ee
induces a pairing of $\Cliff(k)$ and $\Cliff(l)$ modules giving 
\be
M_\star=\bigoplus_k M_k
\ee
a $\Z$-graded ring structure. Multiplication by the class of an irreducible module in $M_8$ gives an isomorphism $M_{k}\cong M_{k+8}$. Let us define $A_k$ as the cokernal of the homomorphism  $M_{k+1}\rightarrow M_k$ induced by the inclusion $\Cliff(k)\hookrightarrow \Cliff(k+1)$. Then $A_\star =\bigoplus A_k$ is an ideal in $M_\star$, with an  inherited  ring structure, and we have an isomorphism with the reduced real $K$-theory,
\be
A_k\cong \widetilde{KO}(S^k)
\ee 
where the Bott periodicity of $A_k$ essentially follows from that of $\Cliff(k)$ via $M_{k}\cong M_{k+8}$. 
When we discuss the classification of topological phases, we will again appeal to  $K$-theory. To set the scene we must first return to the subject of Majorana fermions.

\section{Majorana in condensed matter}\label{MCM}

While the jury is still out on whether or not Majorana fermions will be discovered in the context of fundamental particle physics, condensed matter systems allow us to elaborate on what Nature naively offers, presenting new opportunities to realise physically Majorana's insight. The fundamental constituents of condensed matter systems are invariably electrons, which carry charge and have distinct antiparticles. Clearly, if Majorana fermions are to be found here they must arise as emergent excitations.

Superconductors, in particular, represent a promising setting for such emergent behaviour since the gauge symmetry associated to charge conservation is broken. Indeed, the prospect  of observing Majorana quasiparticles in midgap excitations of a chiral $p$-wave superconductor has a reasonably long history \cite{kopnin1991mutual,Volovik:1999eh,senthil2000quasiparticle, Read:1999fn,kitaev2001unpaired, Huse:2001,sarma2006proposal}, with early indications  appearing in the particle physics literature  more than 30 years ago \cite{jackiw1981zero}.  Much of the focus has  been on the idea of Majorana zero-modes (Mzm),  also dubbed Majorinos  \cite{Wilczek:2014lwa}, appearing in topologically non-trivial phases of matter. Such quanta are Majorana in the sense that they are their own antiparticles, but the term ``fermion'' is intentionally avoided here because they generically obey non-Abelian anyonic exchange statistics. Their topological character and non-Abelian statistics  make Mzm particularly exciting from the perspective of quantum computing, motivating  substantial experimental and theoretical efforts. We will come to this subject in   \autoref{mzm}, but first we  will briefly  review the emergence of Majorana fermions,  in the original sense of the term, in ordinary (non-topological) superconductors. Although less enticing with respect to  quantum computing, these  examples are closest in spirit to Majorana's original analysis and there are even  proposals to observe them in the laboratory that directly mirror  experiments in  particle physics \cite{Beenakker:2014}.

\subsection{Majorana fermions in superconductors}\label{scmf}

In spite of the apparent contradiction  Majorana fermions (note the very intentional use of the term fermion here) in ordinary superconductors are actually rather ubiquitous. Although this has been generally appreciated for some time now, it was emphasised  and  treated carefully only rather recently in \cite{Chamon:2010ks}.

In \cite{Wilczek:2014lwa} Wilczek presented a very simple toy model that captures the key features while avoiding the complication of actually having to deal with superconductors and spinors. To help set the scene, we will repeat this example here. First, recall that the mass term of the Majorana Lagrangian \eqref{ml} in terms of the chiral projection $\psi_L$ is given by the unusual form,
\be\label{Mm1}
\mathcal{L}_m \sim \psi_{L}^{*}\gamma_0\psi_{L}^{*} +  \psi_{L}\gamma_0\psi_{L}.
\ee
Again, we emphasise that this Majorana mass term breaks the $\Un(1)$ symmetry $\psi_{L/R}\rightarrow e^{i\theta} \psi_{L/R}$ of the Dirac Lagrangian and is lepton number violating. Now  consider a model with two complex scalar fields, $\phi$ and $\varphi$, with a global $\Un(1)$ symmetry, $\phi\rightarrow e^{i\theta}\phi$ and $\varphi\rightarrow e^{i2\theta}\varphi$. The key point is that the expectation value of $\varphi$ can break the $\Un(1)$ symmetry such that  a Majorana mass term is generated.  Let the mass and cubic interaction terms, which must initially be compatible with the $\Un(1)$ symmetry,  be given by
\be
\mathcal{L} = -m^2 \phi^*\phi - \frac{\lambda}{2} \left(\varphi^* \phi^2 + c.c.\right).
\ee
If $\langle \varphi\rangle=0$ then the global $\Un(1)$ is preserved and the $\phi$ quanta  are a particle-antiparticle pair of equal mass $m$  and opposite charge. If, however, $\langle \varphi\rangle=\Delta\in\R$  the global $\Un(1)$ is broken and we have the very suggestive scalar analog of a Majorana mass term,
\be
\mathcal{L} \rightarrow -(m^2+ \lambda \Delta) (\text{Re} \phi)^2- (m^2- \lambda \Delta) (\text{Im} \phi)^2.
\ee
The quanta of the  \emph{real} fields $\text{Re} \phi$ and $\text{Im} \phi$ are their own antiparticles with definite and distinct masses, $\sqrt{m^2\pm \lambda \Delta}$. Note, if the symmetry breaking is small  it is possible that  the other interactions (not considered above) are more close to being diagonal in the $\phi, \phi^*$ basis.  Then for time scales  less than the   $\phi, \phi^*$ oscillation time $\sim m/\lambda \Delta>>1$ the Majorana nature of the system will be hidden for practical purposes.  This serves to highlight the point that even if particles are their own antiparticles they can to  arbitrary accuracy behave as if they are not \cite{Wilczek:2014lwa}, a fact we had in mind when discussing neutrinos. 

With this illustrative example in hand we now turn to the emergence of Majorana fermions in superconductors. Our starting point is the Bogoliubov-de Gennes (BdG) formalism for superconductivity with spatial inhomogeneity. See for example \cite{de1999superconductivity}. The mean-field BdG Hamiltonian is given by,
\be\label{BdG}
\mathcal{H}=\int d^3x \left(\psi^{\dagger} h \psi + \frac{i}{2}(\Delta \psi^{\dagger} \sigma_y \psi^{\dagger} + h.c.) -\frac{1}{V}|\Delta|^2  \right),
\ee
Here $\psi^{\dagger}_{s}(\mathbf{x}), s=\uparrow, \downarrow $ creates an electron of spin $s$ at spatial point $\mathbf{x}$. The first term $\psi^{\dagger} h \psi$ gives the kinetic and single-electron potential energy.   The remaining terms are derived from the Bogoliubov mean-field decoupling of the standard attractive  $(V>0)$ interaction term, 
\be
V\psi^{\dagger}_{\uparrow} \psi^{\dagger}_{\downarrow}\psi_{\uparrow} \psi_{\downarrow} \quad \longrightarrow \quad V\left( \langle\psi^{\dagger}_{\uparrow} \psi^{\dagger}_{\downarrow} \rangle\psi_{\uparrow} \psi_{\downarrow}+ \psi^{\dagger}_{\uparrow} \psi^{\dagger}_{\downarrow} \langle\psi_{\uparrow} \psi_{\downarrow}\rangle-\langle\psi^{\dagger}_{\uparrow} \psi^{\dagger}_{\downarrow} \rangle\langle\psi_{\uparrow} \psi_{\downarrow}\rangle\right),
\ee
where we have defined the superconducting  order parameter,
\be\label{sco}
\Delta(\mathbf{x}) = V\langle\psi_{\uparrow} \psi_{\downarrow}\rangle,
\ee
and the expectation values are taken with respect to \eqref{BdG}.
Note that the resulting interaction given by the middle term of \eqref{BdG} for constant $\Delta$ is very much akin to  the conventional number-violating Majorana mass term, as in \eqref{Mm1}, with Majorana mass $|\Delta|$. This is number violating ($\delta L =2$) in the sense that a pair of electrons can be absorbed/produced by the superconducting condensate treated in  the mean-field approximation. These processes have observable consequences, produced by pair-wise annihilation of Bogoliubov quasiparticles, that parallel those of Majorana fermions in particle physics \cite{Beenakker:2014}. 

We can recast \eqref{BdG} in the standard BdG form by introducing the Nambu spinor,
\be
\Psi=\begin{pmatrix} \psi_{s}\\\psi^{\dagger}_{s} \end{pmatrix}
\ee
such that 
\be\label{BdG2}
\mathcal{H}=\int d^3x \left(\Psi^{\dagger} H \Psi -\frac{1}{V}|\Delta|^2  \right),
\ee
where
\be
 H= \begin{pmatrix}h& \Delta \\ \Delta^* & -\sigma_y h^* \sigma_y \end{pmatrix}.
\ee
This form brings us even closer to the original Majorana analysis, since the Nambu spinor is constrained to satisfy the reality condition,
\be
C\Psi^*=\Psi, \quad \text{where}\quad C=\sigma_y\otimes\sigma_y.
\ee
Moreover, expressed in terms of the eigenfuctions of the stationary BdG equation,
\be
H\Psi_n = E_n \Psi_n,
\ee
 used to solve \eqref{BdG2} with the self-consistency condition \eqref{sco}, the BdG Hamiltonian  is exactly of the Majorana Hamiltonian form,
 \be\label{BdG3}
\mathcal{H}=\sum_{E_n>0} E_n  a^{\dagger}_{n}a_n+\text{const},
\quad\text{
 where }
 \quad
 a_n=\int d^dx\Psi^{\dagger}_{n}\Psi_n.
 \ee
 As a result the quasiparticles of the BdG formalism have all the key features of Majorana fermions: they are neutral fermions that are their own antiparticles with number-violating Majorana mass given by the superconducting order $\Delta$. 

As emphasised in \cite{Chamon:2010ks} these Majorana features are a generic consequence of superconductivity and fermionic statistics.   Fermionic statistics alone, without additional assumptions regarding  any other symmetries of the system, very broadly   implies that the BdG Hamiltonian allows real-valued solutions for the Nambu spinor that can be quantised as Majorana fields. They proceed by introducing a generic set  of fermionic degrees of freedom, $\Psi_a$, where $a=(\mathbf{x}, \alpha, s, i)$ is a composite  index labelling position $\mathbf{x}$, possible flavours $\alpha$, any half-integer spin $s$ and Nambu grading $i=\pm$, i.e.~$\Psi_{+(-)}=\psi^{(\dagger)}$. The Hamiltonian density describing superconductivity for such a system can then be written as,
\be\label{BdG4}
\Psi^\dagger H_{\text{gen}}\Psi, \quad\text{where}\quad H_{\text{gen}}=\begin{pmatrix}h& \Delta \\ \Delta^\dagger & - h^*  \end{pmatrix}
\ee
and $h=h^\dagger, \Delta^T=-\Delta$. For consistency, the operators $\Psi$ must satisfy the constraint 
\be
\Psi^{\dagger}=C\Psi, \quad\text{where}\quad C=\begin{pmatrix}0& \mathds{1} \\ \mathds{1} & 0  \end{pmatrix}
\ee
and the anticommutation relations are given by,
\be
\{\Psi_a, \Psi_b\}=C_{ab}, \qquad \{\Psi_a, \Psi^{\dagger}_{b}\}=\delta_{ab}.
\ee
The fermionic nature of $\Psi$ implies that any BdG-type $H_{\text{gen}}$ has  a conjugation symmetry of the form,
\be\label{cc}
-H_{\text{gen}}^*=C^*H_{\text{gen}}C,
\ee
where  $C^*=C^{-1}$. The point highlighted by \cite{Chamon:2010ks} is that there is a unitarily related basis in which $H_{\text{gen}}$ is pure imaginary. Consider a unitary $V$ under which 
\be
H_{\text{gen}}\rightarrow \tilde{H}_{\text{gen}}=VH_{\text{gen}}V^\dagger.
\ee
From \eqref{cc} we have,
\be
 \tilde{C}=VCV^T,
\ee
which respects $ \tilde{C}^*= \tilde{C}^{-1}$. It was shown in \cite{Chamon:2010ks} that for generic $H_{\text{gen}}$ there exists a $V$ such that $C=\mathds{1}$. Then from \eqref{cc} we have $\tilde{H}_{\text{gen}}^*=-\tilde{H}_{\text{gen}}$ so that Schr\"odinger's equation,
\be
\partial_t \tilde{\Psi} = i\tilde{H}_{\text{gen}}\tilde{\Psi} 
\ee
 admits real solutions, just as in Majorana's original analysis. Since the BdG formalism, which builds on the standard Bardeen, Cooper and Schrieffer model, is experimentally very successful, we are led to infer that Majorana fermions in this sense are common features of superconductors. Majorana zero-modes associated to localised states in topologically non-trivial phases are much rarer however. Let us now turn to this topic. 

\subsection{Majorana zero-modes}\label{mzm}

An important theme of current research in condensed matter is the notion of topological phases. These are not distinguished in the familiar manner by  symmetries, but instead by topological invariants. They now come in a variety of guises, several of which have be experimentally realised \cite{Kane:2005, Bernevig:2006, Moore:2007, Konig:2007, Fu:2007, Hsieh:2008fk,Roy:2009,Hsieh:2009, Chen:2009}. Typically topological insulators or superconductors are fermion systems that have  a gap in the bulk, but admit topologically protected gapless boundary states.  

The theoretical and experimental existence of such states has dramatic implications, both in terms of fundamental physics and  potential applications: they represent a genuinely new feature of quantum theory while offering an alternative  path to fault-tolerant  quantum computation. The susceptibility of qubit systems to decoherence, leading to phase and spin-flip errors, remains the principal obstacle to  realising a scalable quantum computer.  One approach is to use  fault-tolerant  error-correcting  protocols, algorithmically redundant portions of code that serve specifically to detect and correct decoherence errors \cite{Aharonov:1997,Knill:1996}. If the error occurrence rate can  be kept under a threshold of around $10^{-4}$ \cite{Zalka:1996nc, Steane:1997}, then fault-tolerant  error-correction would allow for large scale  reliable quantum computation.  Topological quantum computing \cite{Kitaev:1997wr} is another approach to fault tolerance, complementary to the detect and correct paradigm. The idea is to use physical features of the system, specifically topological obstructions, to protect against errors appearing  in the first place. Majorana zero-modes arising in  topologically non-trivial  superconducting  phases  are a striking  example of how these ideas may be realised in practise. 

In \autoref{mc} we will discuss the appearance of  Mzm in a paradigmatic example, the  Majorana (or Kitaev) chain \cite{Kitaev:2001}.  We then briefly consider there application to topological quantum computing in \autoref{qc}.  To conclude, we will address the definition and classification of topological phases for gapped free-fermion systems more carefully and generally in \autoref{topclass}. 

\subsubsection{The Majorana chain}\label{mc}

 The simplest model admitting Mzm is the 1-dimensional Majorana (or Kitaev) chain \cite{Kitaev:2001}. A length $N$ Majorana chain consists of $N$ spinless  fermions on a 1-dimensional lattice. The spinless condition  initially gave the impression that this example would be  limited to an unrealistic, albeit illuminating, toy model.   However, there are now several plausible physical realisations, as we shall briefly review later. 

Before turning  to Kitaev's model, let us discuss some generic features and expectations.  The basic idea is that each fermionic site $i=1,\ldots N$ is described by the usual fermionic annihilation and creation operators, $a_i, a_{i}^{\dagger}$, with the usual anti-commutation relations, $\{a_{i}^{\dagger}, a_{j}\}=\delta_{ij}, \{a_{i}^{(\dagger)}, a_{j}^{(\dagger)}\}=0$. Here, $a_{i}^{\dagger}$ and $a_i$ can be thought of as creating a ``spinless electron'' or hole, respectively, at site $i$. These can be formally arranged into a set of so-called Majorana operators, 
\be
\gamma_{2i-1}=a_{i}^{\dagger}+a_i,\qquad \gamma_{2i}=i(a_{i}^{\dagger}-a_i)
\ee   
which satisfy the Clifford and Majorana like relations
\be
\{\gamma_{\alpha}, \gamma_{\beta}\} = 2\delta_{\alpha\beta},\qquad  \gamma_{\alpha}^{\dagger}=\gamma_{\alpha}, \qquad \alpha= 1,\ldots 2N.
\ee
These modes are their own antiparticles in the sense that $\gamma^{\dagger}=\gamma$, but note that $\gamma^2=1$ and  therefore $\gamma$ is not fermionic in  conventional terms. Generically this is of course just a formal manipulation and we  should not think of the $\gamma$ modes as representing quasiparticles. It is the $a_i=\gamma_{2i-1}+i\gamma_{2i}$ operators that typically have a well-defined occupation number; the Majorana modes  pair up into ordinary ``Dirac'' modes with distinct antiparticles, $a^\dagger\not=a$. 

However, Kitaev demonstrated \cite{Kitaev:2001} that there is a  natural family Hamiltonians that give rise to Majorana modes as the effective low-energy degrees of freedom. The key observation is that, in special circumstances, two   single Majorana modes can be localised at the boundaries of the chain, rendering them effectively free and   un-paired  \cite{Kitaev:2001}. For such systems the Majorana operators are not merely a formal tool, but are actually required to capture the correct physics. 

Kitaev considered the Hamiltonian,
\be\label{kc}
\mathcal{H}=\sum_i \left[-t (a^{\dagger}_{i} a_{i+1}+h.c.)-\mu(a^{\dagger}_{i} a_{i}-\frac{1}{2})+(\Delta a_{i} a_{i+1}+h.c.)\right],
\ee
where $t$ is the hopping amplitude, $\mu$ is the chemical potential and $\Delta=e^{i\theta}|\Delta|$ the induced superconductiing gap. For $\Delta\not=0$ this represents a superconducting quantum wire of spinless fermions. 

Ignoring the phase $\theta=0$ (via the obvious gauge transformation) for simplicity we have 
\be
\mathcal{H}=\frac{i}{2}\sum_i \left[\mu \gamma_{2i} \gamma_{2i-1}+(\Delta+t) \gamma_{2i} \gamma_{2i+1}+ (\Delta-t) \gamma_{2i-1} \gamma_{2i+2}\right]
\ee
in terms of the Majorana operators.
Kitaev showed that this Hamiltonian admits  two phases divided by the lines $2|t|=|\mu|$:
\begin{itemize}
\item[$(a)$]
 $2|t|<|\mu|$, topologically trivial with all pairs of Majorana modes belonging to a common site $i$ bound  into an ordinary ``Dirac" fermion.
 \item[$(b)$] $2|t|>|\mu|, \Delta\not=0$, topologically non-trivial with  un-paired Majorana zero-modes appearing at the ends of the chain. 
\end{itemize}
To understand these two phases it is instructive to consider a simple limiting example from each class. 

For $(a)$ let $t=\Delta=0, \mu<0$ so that 
\be
\mathcal{H}=\frac{i}{2}(-\mu)\sum_i   \gamma_{2i-1} \gamma_{2i}=-\mu \sum_i \left(a^{\dagger}_{i} a_{i}-\frac{1}{2}\right).
\ee
We immediately see that  the Majorana operators $ \gamma_{2i-1}, \gamma_{2i}$ associated to a given site $i$ are paired together and the ground state has all sites unoccupied. The system is obviously gapped since it costs energy $\mu$ to excite a quasiparticle. 

For $(b)$ let $t=\Delta>0, \mu=0$ so that 
\be\label{H1}
\mathcal{H}=it\sum_{i}^{N-1}  \gamma_{2i} \gamma_{2i+1}=2t \sum_{i}^{N-1} \left(c^{\dagger}_{i} c_{i}-\frac{1}{2}\right),
\ee
where
\be
c_{i}=\frac{1}{2}(\gamma_{2i}+i\gamma_{2i+1}),\quad c^{\dagger}_{i}=\frac{1}{2}(\gamma_{2i}-i\gamma_{2i+1}).
\ee
We see that in the bulk  Majorana operators $ \gamma_{2i}, \gamma_{2i+1}$ associated to adjacent but \emph{different}  sites, $i$ and $i+1$, are paired together.  The extremal  operators $\gamma_1=\gamma_L$ and $\gamma_{2N}=\gamma_R$ are Majorana zero-modes; they do not appear in the Hamiltonian at all and are    unpaired.  From the second form of \eqref{H1} it is apperent that the ground states satisfies $c_i|\Omega_0\rangle=0$ for all $i<N$. Since $\gamma_{L,R}$ obviously commute with the Hamiltonian there is a two-fold degeneracy of the ground state.   In particular, the  non-local ``Dirac'' fermion 
\be
f=\frac{1}{2}(\gamma_L+i\gamma_R),
\ee 
costs zero energy and if $f|\Omega_0\rangle=0$ then $|\Omega_1\rangle=f^\dagger|\Omega_0\rangle$ is necessarily a second orthogonal ground state. We can  write,
\be\label{gs}
-i\gamma_L \gamma_R|\Omega_0\rangle=|\Omega_0\rangle,\qquad -i\gamma_L \gamma_R|\Omega_1\rangle=-|\Omega_1\rangle
\ee
from which we see that $|\Omega_0\rangle$ is even and  $|\Omega_1\rangle$ is odd 
under  the fermionic parity operator
\be\label{parity}
P=\prod_i(-i\gamma_{2i-1}\gamma_{2i}).
\ee
As for $t=\Delta=0, \mu<0$ the bulk has a gapped spectrum since it cost $t$ to produce an excitation. In fact, the bulk properties of the two cases are identical. However, the boundary conditions remain distinct; only for $t=\Delta>0, \mu=0$ can unpaired Majorana zero-modes exists. In this sense $t=\Delta>0, \mu=0$ belongs to a topologically non-trivial phase. We will make this statement more precise in \autoref{topclass}. 

In summary, these two cases represent two distinct phases of the Majorana chain that have the same bulk features, but differing boundary modes. Only in the latter are there unpaired zero energy Majorana modes. The properties of these  two phases extend to the full parameter spaces defined in $(a)$ and $(b)$ above. This relies on the fact the bulk spectrum is gapped in both phases. For a finite length chain this suppresses  exponentially in $N$ the amplitude for the fermionic pair described by $\gamma_{L,R}$ to tunnel across the chain. To study the bulk energy spectrum for generic values of the parameters we can impose periodic boundary conditions. The Hamiltonian in momentum space is then given by 
\be\label{mom}
\mathcal{H}=\sum_k \left[(-2t \cos(k) -\mu)a_{k}^{\dagger}a_k+\Delta (i\sin(k) a_ka_{-k}+h.c.) \right]
\ee 
and the energy is given by 
\be
E(k)=\pm\sqrt{(2t \cos(k)+ \mu)^2 +(\Delta \sin(k))^2}
\ee
so that for $\Delta\not=0$ the spectrum is fully gapped unless $2t=\pm\mu$. The key condition for the two phases to extend to the full parameter spaces $(a)$ and $(b)$ is the absence and existence, respectively, of zero-energy boundary modes linear in the Majorana operators. That is, solutions to the equation
\be
[\mathcal{H}, \sum_{\alpha=1}^{2N} c_\alpha \gamma_\alpha] =0. 
\ee
A very clear treatment of this problem is given in \cite{Wilczek:2014lwa}, so we will not repeat it here. Instead, let us summarise the key attributes of the solution. As one would expect, it splits into two cases. For the parameter range specified by $(a)$ the required boundary conditions (for an open chain) cannot be satisfied and hence there are no zero-mode solutions. If, on the other hand, $2t > |\mu|, \Delta\not= 0$ then all boundary conditions are satisfied (up to corrections exponentially suppressed in $N$) and we have two boundary zero-modes $b_L$ and $b_R$ localised at $i=1$ and $i=N$, respectively. If $2t > - |\mu|, \Delta\not= 0$ then $b_L$ and $b_R$ are simply interchanged and we conclude that boundary Majorana zero-modes exist for the whole of the $(b)$ parameter range. 

For $N\rightarrow\infty$ this result is exact. For finite $N$ there is a weak interaction between $b_L$ and $b_R$ that can accounted for with an effective Hamiltonian, 
\be
\mathcal{H}_{\text{int}}\propto e^{-N/l_0} b_L b_R.
\ee 
Here $l_{0}^{-1}$ is the characteristic length scale given by the smallest of $|\ln |\mu_+||$ and $|\ln |\mu_-||$, where
\be
\mu_\pm = \frac{-\mu\pm\sqrt{\mu^2-4(\Delta^2-t^2)}}{2(\Delta+t)}.
\ee
Note, for  $(b)$ we have  $|\mu_\pm|<1$. The interaction term  drives a  re-combination of  $b_L$ and  $b_R$ back into a conventional doublet of creation and annihilation operators.  However, for a long wire  the operators $b_L, b_R$ can be effectively regarded as separated and local up to exponentially suppressed corrections. The degeneracy of the grounds states is lifted  by a separation of $e^{-N/l_0}$ and up to exponentially suppressed corrects the approximate Majorana zero-modes satisfy the characteristic equation $b_{L,R}^{2}=1$.  Note, we have neglected the treatment of (weak) interaction terms beyond the mean-field approximation and the possible effect they may have on the existence of Majorana modes, although they are expected to persist. For a recent discussion of these issues see \cite{Kells:2015, Kells:2015a} and the references therein.

Note, the above analysis of Mzm rested on the assumption that the chain is made up of spinless fermions. Were it otherwise, the degeneracy of every eigenstate would be doubled and, in particular, the boundaries would support two Mzm  re-assembled into a single ordinary fermion. That is not to say the Kitaev chain cannot be realised using electron systems, only that  their spin degree of freedom must be essentially frozen out. There are now a number of rather ingenious and realistic proposals that address this issue, while meeting the other two key requirements of superconductivity and a bulk gap. These include the boundary of  2-dimensional topological insulators   \cite{PhysRevB.79.161408, RevModPhys.82.3045, RevModPhys.83.1057, franz2013topological}  or nanowires made from a  3-dimensional topological insulator  \cite{PhysRevB.84.201105}, both with proximity induced superconductivity. 

The experimentally most successful approach to date is the use of semiconducting quantum wires. The pioneering proposals of \cite{PhysRevLett.105.077001,PhysRevLett.105.177002} use a semiconducting wire  placed on the surface of a block of conventional 3-dimensional  $s$-wave superconductor with an external magnetic field applied. If the wire has an appreciable spin-orbit coupling then the magnetic field can be used to prefer a particular spin direction, effectively rendering the constituent electrons spinless.  Remarkably, the proximity effects of the $s$-wave superconductor, together with the spin-orbit coupling, can effectively induce   $p$-wave superconductivity in the wire as required by the Cooper-pairing of spinless fermions. For a  review of this phenomenon see  \cite{alicea2012new}. The freedom to use readily available   $s$-wave superconductors is obviously of great technological significance. 

The first experimental evidence supporting the existence of Mzm was observed in the breakthrough work of the Delft group \cite{Mourik1003}. They placed a single InSb crystal wire, which has good proximity induced superconductivity, on a substrate. One portion of the wire is placed in contact with a superconducting metal and  another disconnected portion is placed in contact with a normal metal. The small middle section of the wire not in contact with either is depleted of electrons to create an approximately insulating bridge connecting the two portions. Theoretically, with an appropriate  external $B$ field applied we would expect the superconducting portion to enter a topological phase with Mzm localised at the ends.  A voltage bias is applied across the bridge and the differential tunnelling conductance is measured. To a very good approximation the conductance is proportional to the density of states at the  end of the superconducting portion of the wire adjacent to the bridge.  For a small external magnetic field, less than about 90 mT, the observed conductance suggests a superconducting gap at around 260 $\mu$eV and nothing special happening at zero bias, consistent with a wire in the expected topologically trivial phase. As the $B$ field is increased a peak in the conductance is seen  at zero bias suggesting a transition to a topological phase and the emergence of Mzm qualitatively consistent with the theoretical prediction. This result has since been reproduced by a number of groups \cite{PhysRevB.87.241401,das2012zero, doi:10.1021/nl303758w, PhysRevLett.110.126406}, amounting to a compelling case for the existence of Mzm. It should be noted, however, that a similar zero-voltage bias conductance peak can be induced by  disorder and multiple bands even when in a topologically trivial phase \cite{PhysRevLett.109.267002}, so perhaps some caution is still required.

\subsubsection{Topological quantum computing}\label{qc}

As we have emphasised several times Mzm are not fermions in the conventional sense. Instead they are \emph{anyons}, a term coined by Wilzcek \cite{PhysRevLett.49.957}; identical particles with non-trivial quantum exchange statistics that is neither  bosonic nor fermionic. The possibility of anyonic statistics was first demonstrated in \cite{Leinaas2007, PhysRevLett.48.1144}. Note, in three space dimensions quantum statistics is necessarily  either bosonic or fermionic.  Adiabatically exchanging a pair of particles twice is equivalent to adiabatically sending one particle all the way around the other, which in three space dimensions is topologically equivalent to having done nothing and thus must act on the wavefunction as the identity. The exchange operator squares to the identity implying  one of the two standard possibilities: boson versus fermion. In two space dimensions, however, this is no longer true \cite{Leinaas2007, PhysRevLett.48.1144}; a path enclosing a particle cannot be smoothly deformed to a point without being cut by the particle.    When a pair of \emph{Abelian} anyons  are interchanged  the wavefunction picks   up an  complex phase factor and, as such, they can be thought of  as interpolating between bosons and fermions.  Abelian ayons are now recognised as a ubiquitous feature of fractional quantum hall states \cite{PhysRevLett.52.1583, PhysRevLett.53.722}. 

More recently, it has been realised that there are ayonic system for which particle exchange induces a unitary change of state going  beyond a mere  phase factor. In this case the ordering of exchanges is important and they are accordingly referred to as \emph{non-Abelian} anyons. Quasiparticles with non-Abelian aynionic statistics first arose in conformal field theory \cite{Moore1988451} and Chern-Simons theory \cite{witten1989}.   Later, Moore and Read \cite{Moore1991362} showed that  fractional quantum Hall states could support non-Abelian statistics. There is now significant  theoretical and experimental evidence for non-Abelian statistics in quantum Hall states with filling factor 5/2 \cite{dolev2008observation, Radu899, Willett02062009}.  A key step with respect to our present story  was the discovery \cite{PhysRevB.61.10267} that these Moore-Read states share universal features, including non-Abelian anyonic statistics,  with those of topological 2-dimensional spinless $p+ip$ superconductors. In groundbreaking work, Fu and Kane \cite{PhysRevLett.100.096407} argued that  the required topological phase in two dimensions would emerge at the interface of a topological insulator and a conventional $s$-wave superconductor.  Mzm that commute with the Hamiltonian  are localised at the centres of Abrikosov vortices\footnote{For our purposes it is sufficient to regard a vortex as a point at which the superconducting order $\Delta$ goes to zero and around which its phase picks up a $2\pi$ shift.} 
 that form in the  topological superconducting interface and are subject to non-Abelian exchange statistics. 
Experimental realisations of the Fu-Kane model are in progress \cite{PhysRevLett.109.056803, cho2013symmetry,zareapour2012proximity,wang2013fully} and there have  been some early indications of Mzm reported in \cite{PhysRevLett.112.217001}.

Such systems, realising non-Abelian Mzm in two dimensions, represent a compelling framework for topological quantum computing, as first proposed by Kitaev \cite{Kitaev:1997wr, freedman1998p, Kitaev:2001, kitaev2006anyons}. Unfortunately, we cannot even scratch the surface here. Reference  \cite{RevModPhys.80.1083} is  recommended for the interested reader. Consider a distribution of $2N$ Mzm $\gamma_i, i=1,2,\ldots,2N$ (localised by vortices) arranged in a 2-dimensional plane and protected by a gap. As in the Kitaev chain, the Mzm are paired into $N$ non-local fermion operators $f_i=(\gamma_{2i-1}+i\gamma_{2i})/2$, which generate $2^N$ orthogonal zero-energy states that can be either occupied or empty, $n_i=0,1$. The subspace of ground states $\mathcal{H}_0$ is thus $2^N$-dimensional and spanned by,
\be\label{basis}
|n_1n_2,\ldots n_N\rangle, \qquad n_i\in\{0,1\},
\ee
where the $n_i$ are the eigenvalues of the corresponding number operators,
\be
n_i=f_{i}^{\dagger}f_i=\frac{1}{2}\left( 1+i\gamma_{2i-1}\gamma_{2i}\right).
\ee
The ground state manifold of $2N$ Mzm thus renders us a basis for $N$ topological qubits \cite{nayak19962n, Kitaev:1997wr}. The inherent non-locality of the $f_i$ operators dramatically suppresses environmentally induced decoherence, since it requires coordinated  ``measurements'' at spatially separated locations. 

\begin{figure}
\centering
\includegraphics[scale=0.5]{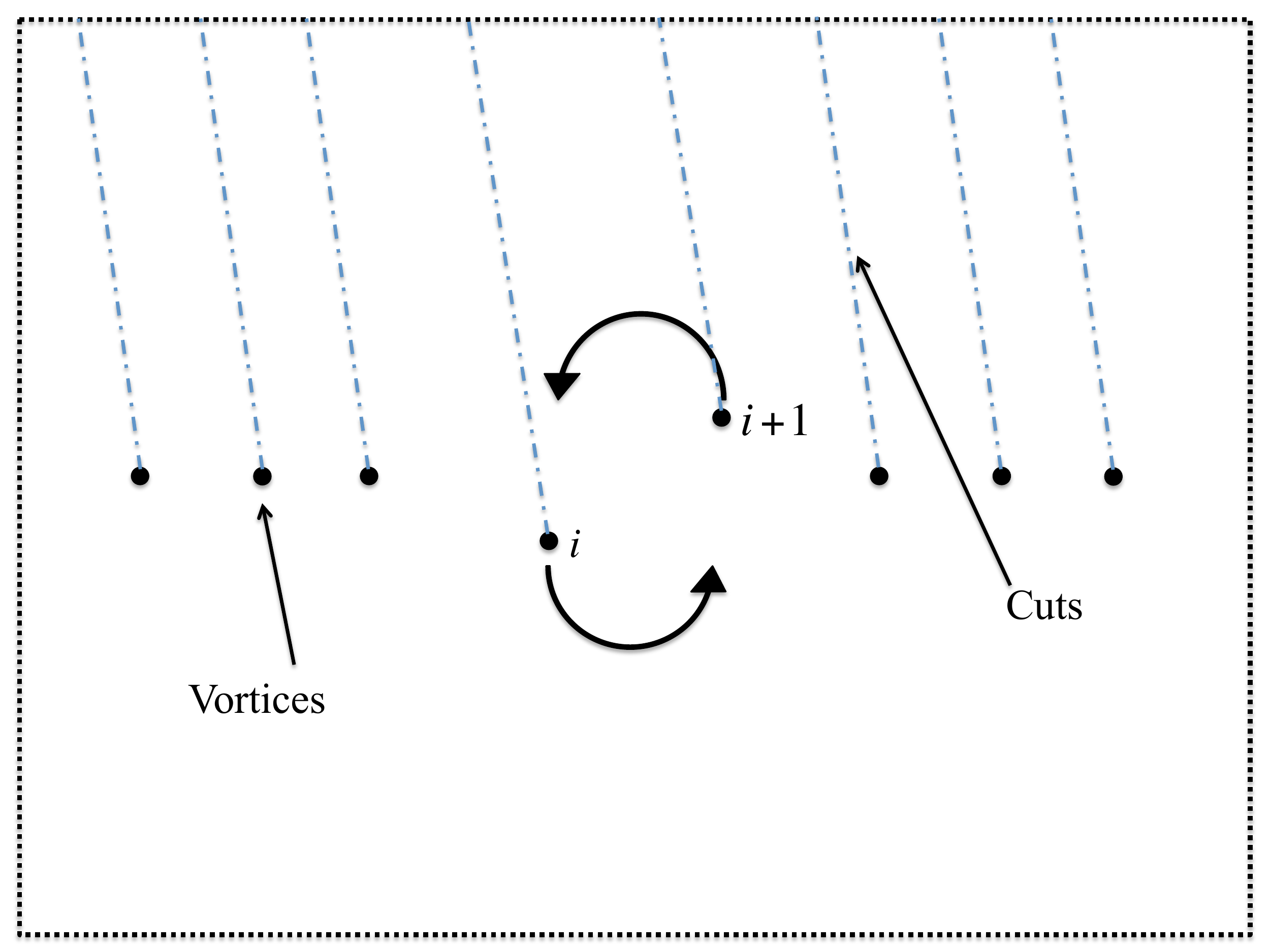}
\caption{Exchanging vortices.\label{vort}}
\end{figure}

If the temperature is kept sufficiently below the bulk gap then an initial qubit state will remain  confined to the ground state manifold under the adiabitic manipulation of the vortices  and, hence, the  state space explored  is protected against local perturbations. However, because of the non-Abelian  statistics the adiabatic exchange of two vortices will generically induce a unitary transformation within the qubit state space.  Not only do the topological features of the system protect against decoherence errors, a large class of unitary operations acting on the qubit state can be implemented by simple sequences of adiabatic swaps!

Let us explore this last idea a little further. The non-Abelian nature  of Mzm in the Fu-Kane model (and its ilk)  essentially follows from  geometric phase and explicit monodromy  considerations, see for example \cite{gurarie1997plasma, Moore1991362, Read:1999fn,  stern2004geometric}.  Ivanov \cite{PhysRevLett.86.268}  developed a  heuristically useful picture  that reduces the essence of the phenomenon to a set of rules for its effect on the Mzm operators $\gamma_i$ as the vortices are interchanged. Consider a set of vortices in the plane as in \autoref{vort}. We let the vortices adiabatically move around before returning to the same set of positions, but possibly re-ordered. To encode the phases picked up Ivanov attached a cut extending from each vortex to a common boundary. Each time a vortex passes through a cut the corresponding Mzm operator acquires a phase shift of $\pm\pi$ and so picks up a sign.  Roughly speaking this is related to the phase  of the superconducting order parameter $\Delta$, which is regarded as single-valued away from the cuts and jumps by $2\pi$ across the cuts. From \eqref{BdG} a jump of $2\pi$ corresponds to a phase shift of $\pi$ for the Mzm operators.  From this rule  we see that if two adjacent vortices $i$ and $i+1$, as in \autoref{vort},  are interchanged in a counterclockwise fashion (without interfering with any other vortices) we obtain the following transformation:
\be
T_{i}:
\left\{\begin{array}{llllll}
\gamma_i&\mapsto& \gamma_{i+1}\\
\gamma_{i+1}&\mapsto& -\gamma_{i}\\
\gamma_{j}&\mapsto& \gamma_{j}, \quad \forall j\not=i, i+1\\
\end{array}\right.
\ee
Here we are using $T_i$ to denote the operation induced by counterclockwise exchanging the  vortex at postition  $i$ with vortex at position $i+1$. The operation induced by the vortex at position $i$ encircling the vortex at position $i+1$ is therefore given by 
\be
T_{i}^{2}:
\left\{\begin{array}{llllll}
\gamma_i&\mapsto& -\gamma_{i}\\
\gamma_{i+1}&\mapsto& -\gamma_{i+1}\\
\gamma_{j}&\mapsto& \gamma_{j}, \quad \forall j\not=i, i+1\\
\end{array}\right.
\ee
consistent with the observation that each vortex must cross the others cut once. Note, it is also simple to check  that these rules imply $T_{i}^{4}=\mathds{1}$ and the relations:
\be\label{br}
\begin{array}{ccclll}
T_iT_j&=&T_jT_i, & |i-j|>1\\[5pt]
T_iT_jT_i&=&T_jT_iT_j, & |i-j|=1.
\end{array}
\ee
These are the relations obeyed by the braid group,
\be
B_{n} = \langle ~T_1,\ldots, T_{n-1} ~|~ T_iT_j=T_jT_i\quad |i-j|>1\quad \text{and}\quad T_iT_jT_i=T_jT_iT_j \quad  |i-j|=1~\rangle.
\ee
Indeed, if we adopt a vortex  world-line  point of view, sequences of oriented exchanges  are obviously mapped one-to-one to braids. The braid group is generated by the elementary exchanges $T_i$ modulo the relations \eqref{br}, where composition is given by sequential application of the braid operations.  The action of the braid group generators on the Mzm operators induces a projective representation $\tau:\mathcal{H}_0\rightarrow\mathcal{H}_0$ acting on the Hilbert space of ground states. Since the basis \eqref{basis} can be constructed using the the Mzm operators, the generating elements of $B_{2N}$ take a simple explicit form in terms of Mzm operators \cite{PhysRevLett.86.268},
\be
\tau(T_i) = \exp(\frac{\pi}{4} \gamma_{i+1}\gamma_i)=\frac{1}{\sqrt{2}}(1+\gamma_{i+1}\gamma_i).
\ee
This is essentially the Jones representation of the braid group \cite{Jones:1987}, which is reducible for an even number of strands. Indeed, since the generating operators or even in Mzm operators they preserve the fermion number mod 2 and the Hilbert space has two $2^{N-1}$-dimensional  invariant subspaces. This same representation was found previously by Nayak and Wilczek \cite{nayak19962n} using conformal field theory. As an example consider the action on the 2-qubit (four Mzm) basis vectors of $T_{2}^{-1}$ (\emph{clockwise} exchange of vortex two and three),
\be
\tau(T_{2}^{-1}) |n_1n_2\rangle = \exp(-\frac{\pi}{4} \gamma_{3}\gamma_2) |n_1n_2\rangle = \frac{\pi}{4}\left(|n_1n_2\rangle +i(-1)^{n_1} \sigma_x\otimes\sigma_x |n_1n_2\rangle\right).
\ee

Given the importance of two dimensions here it would seem naively that the 1-dimensional Kitaev chain is not of use;  two quasiparticles necessarily have to pass through each other  when exchanged in one dimension, thus overlapping and breaking the ground state degeneracy. Indeed, Kitaev originally envisaged  the chain as a form of topologically stable quantum memory for this reason. However, this obstacle is circumvented by fabricating 2-dimensional, or even 3-dimensional, networks of quantum wires \cite{alicea2011non,PhysRevB.84.094505, halperin2012adiabatic}. Through an ingenious mechanism the Mzm at junctions of the network can be adiabatically swapped using locally tuneable voltage   gates that shift the topological domains along the wires \cite{alicea2011non,halperin2012adiabatic}. The simplest example is the three-point move at a T-junction, which allows the two possible exchanges of two Mzm.  Since the argument for non-Abelian statistics in topological 2-dimensional spinless $p+ip$ superconductors relied on the exchange of vortices binding the Mzm it is not immediately obvious that the quantum wire networks should exhibit the same behaviour. However, it was shown in \cite{alicea2011non,PhysRevB.84.094505, halperin2012adiabatic} that the effective $p$-wave pairing induced in the wires by the $s$-wave superconductor contacts leads to precisely the behaviour described by Ivanov for Mzm bound to vortices, so that the networks have the same topological quantum computing potential.  A nice pedagogical discussion of the  related algebraic structures that emerge is given in \cite{Wilczek:2014lwa}.

Finally, we should note that the braiding operations of 2-dimensional spinless $p+ip$ superconductors or quantum wire networks alone are not sufficient for universal quantum computing, despite the fact they can implement a significant subset of the required unitary transformations. Fortunately the required additional operations can be implemented non-topologically without paying too  high a price \cite{PhysRevA.71.022316, PhysRevA.73.042313}.   Alternatively, one can search for generalisations of the quantum wire networks for which the braiding moves yield quantum operations rich enough  to generate all the gates required for universal quantum computation \cite{Freedman:2002}.  
 
\subsubsection{The periodic table of topological phases}\label{topclass}

We previously noted that the Majorana chain has two phases, one of which is topologically non-trivial.  We will now make this statement more precise following \cite{Kitaev:2001, PhysRevB.78.195125, budich2013equivalent}. In particular, it will be shown how the two phases are distinguished by a topological invariant. We will then place this example within the general classification of gapped free-fermion systems and their phases  \cite{altland1997nonstandard, PhysRevB.78.195125, Kitaev:2009mg,Ryu:2010,Kennedy:2014cia}.

For the Majorana chain with open boundary conditions the  two phases have the same bulk symmetries and physics, but differ in their edge states. This situation is clearly not captured by the standard Landau paradigm, raising the question of whether or not the phases can be distinguished by studying the bulk physics. Indeed they can using a topological invariant that depends only on the properties of the bulk Hamiltonian. More generally, two gapped phases  are said to be \emph{topologically equivalent} if there is a continuous path in the phase diagram connecting them without closing the gap at any point. Phases that are topologically equivalent to a collection of independent atoms are said to be \emph{topologically trivial}, otherwise they are \emph{topologically non-trivial}. These phases are distinguished not by symmetries, but by \emph{topological invariants} that are insensitive to smooth deformations  of the underlying topological space. These words are, well, just words and there are a number of approaches to making them more mathematically precise, building largely on the pioneering classification schemes of \cite{altland1997nonstandard, PhysRevB.78.195125, Kitaev:2009mg,Ryu:2010}. Later we will comment on this challenge, but first let us return to the treatment of the Majorana chain.

Assuming an energy gap, for each 1-dimensional Hamiltonian $\mathcal{H}$  Kitaev argued  on physical grounds for the existence of a $\Z_2$ valued topological invariant,  the Majorana number $\mathcal{M}(\mathcal{H})\in\{-1,1\}$, that takes fermionic parity into account. Consider two parallel weakly interacting open Majorana chains. The   un-paired Mzm localised on the boundary of each chain may pair-up breaking the ground state degeneracy so that if $\mathcal{M}(\mathcal{H})=-1$ indicates the existence of Mzm then the Majorana number should satisfy $\mathcal{M}(\mathcal{H}\oplus \mathcal{H})=\mathcal{M}(\mathcal{H})\mathcal{M}(\mathcal{H}')$ for non-interacting chains. For a fixed Hamiltonian $\mathcal{\mathcal{H}}$, Kitaev showed that for two  closed chains of lengths $N_1$ and $N_2$ the Majorana number can be expressed as
\be
\mathcal{M}(\mathcal{H})=\frac{P(\mathcal{H}(N_1+N_2))}{P(\mathcal{H}(N_1))P(\mathcal{H}(N_2))}.
\ee
Here $\mathcal{H}(N)$ denotes the Hamiltonian on a closed chain of length $N$ and $P(\mathcal{H}(N))$ is the fermionic parity \eqref{parity} of the unique ground state. This identity follows essentially from considering the two possible ways the two chains  can be closed, either into two separate chains   of length $N_1$ and $N_2$ or a single chain of length $N_1+N_2$.

 Generically, the Majorana number will be difficult to calculate, but for a  gapped  Kitaev chain it can computed quite simply. Using the Majorana operators a completely generic (i.e.~arbitrary, but quadratic) Hamiltonian, subsuming the specific case of \eqref{kc}, for such a system  with $N$ sites can be written, 
\be
\mathcal{H} = \frac{i}{4}\sum_{\alpha,\beta=1}^{2N} \gamma_\alpha A_{\alpha\beta} \gamma_\beta,
\ee
where $A$ is real and antisymmetric. We also assume $A$ is non-degenerate. Consequently there is a $S\in\Orth(2N)$ such that 
\be\label{can}
\tilde{A} = SAS^T = \bigoplus_{i=1}^{N} \lambda_i\varepsilon, \qquad \varepsilon=\begin{pmatrix}0&1\\-1&0\end{pmatrix}.
\ee
is block-diagonal and $\lambda_i>0, \forall i$. Here, $\pm i \lambda_i$ are the pure imaginary eigenvalues of $A$. In this basis the ground state has even parity since it  is annihilated by $\tilde{a}_i=(\tilde{\gamma}_{2i-1}+i\tilde{\gamma}_{2i})/2$, which implies 
\be
\prod_i (-i\tilde{\gamma}_{2i-1}\tilde{\gamma}_{2i}) |0\rangle = \prod_i (\tilde{\gamma}_{2i-1}\tilde{\gamma}_{2i-1}) |0\rangle = |0\rangle.
\ee
For $\det S=1$ or $\det S=-1$  the transformation \eqref{can} is parity preserving or parity reversing, respectively. Hence, on transforming back we obtain,
\be
P(\mathcal{H}) = \text{sgn} \det(S) = \text{sgn}~ \text{Pf}(A), 
\ee
where we have used $\text{Pf}(S{A}S^T)=\text{Pf}(\tilde{A})>0$ and the property of the Pfaffian, $\text{Pf}(S{A}S^T)=\text{Pf}({A})\det(S)$. Note, for an even length chain we therefore have $\mathcal{M}(\mathcal{H})=\text{sgn}~ \text{Pf}(A)$. Since we are dealing with closed chains we have periodicity $N$ and more generally we can consider any periodicity $L|N$. Writing 
\be
\mathcal{H} = \frac{i}{4}\sum_{x,y=1}^{L}\sum_{a,b=1}^{2N/L} \gamma_{xa} A_{ab}(y-x) \gamma_{yb},
\ee
where $y-x$ is taken mod $L$, we can use the Fourier transform 
\be\label{FT}
\bar{A}_{ab}(k) =\sum_{z} e^{ikz}A_{ab}(z), \quad k=\frac{2\pi n}{L}\mod 2\pi\quad\text{for} \quad n\in\Z_N 
\ee
to write the Pfaffian as
\be
\Pf(A)= \prod_{k=-k} \Pf \bar{A}(k) \prod_{k\not=-k} \det \bar{A}(k).
\ee
The assumption that the system is gapped implies that $\det \bar{A}(k)$ is positive \cite{Kitaev:2001} and hence, recalling $k$ is defined mod $2\pi$,
\be\label{pf}
 \mathcal{M(H)}=\text{sgn}~ \text{Pf}(A) = \text{sgn}\left[  \Pf \bar{A}(0) \Pf \bar{A}(\pi) \right].
\ee
For the Majorana chain the momentum space Hamiltonian was given in \eqref{mom}. Arranging the Majorana operators into doublets
\be
\Gamma_i = (\gamma_{2i-1}, \gamma_{2i})^T
\ee
momentum space Hamiltonian \eqref{mom} in the Majorana basis is given by
\be\label{mom2}
\mathcal{H}=i\sum_k \Gamma_{k}^T \left[-(2t \cos(k) -\mu)i\sigma_y+\Delta\sin(k) \sigma_x \right]\Gamma_{-k}
\ee 
so that 
\be
\bar{A}(0) = -(2t -\mu)i\sigma_y, \quad \bar{A}(\pi) =-(-2t -\mu)i\sigma_y
\ee
and
\be
 \text{sgn}~ \text{Pf}(A) = -\text{sgn} (2t -\mu)(2t+\mu)=\begin{cases}\phantom{-}1\quad 2|t|<|\mu|\\-1\quad 2|t|>|\mu| \end{cases}
\ee
in agreement with the analysis of the phases given \autoref{mc}.

The geometric significance  of the Majorana number as given in \eqref{pf} is not immediately apparent. However, there is also a $\Z_2$ valued Chern-Simons invariant defined on the Majorana chain \cite{hatsugai2006quantized} and  it was shown in \cite{budich2013equivalent} that the two forms are equivalent. Using $
\det(S) = \text{sgn}~ \text{Pf}(A)
$
we can alternatively write 
\be
 \mathcal{M(H)}= \ \det \bar{S}(0) \det \bar{S}(\pi) ,
\ee
where the Fourier transform $\bar{S}(k)$ is unitary and  satisfies  $\bar{S}(k)^*=\bar{S}(-k)$ from the reality of $S$. Letting $\det \bar{S}(k)=\exp(i\phi_k)$ we then have 
\be
 \mathcal{M(H)}=(-1)^{\Delta\phi/\pi}, 
\ee
where $\Delta\phi=\phi_0-\phi_\pi$ is quantised in units of $\pi$, since the reality constraint implies $\phi_{k}^*=-\phi_{-k} \mod 2\pi$. In the large $N$ limit the phase difference can be expressed in terms of the Zak-Berry phase,
\be
\Delta\phi=\int^{\pi}_{-\pi} \Tr \mathcal{A}(k)dk
\ee
where $\mathcal{A}(k)$ is the non-Abelian Zak-Berry connection on the set of occupied Bloch states. Hence, the Majorana number is identified as a quantised Chern-Simons invariant in one dimension $\mathcal{M}(H)=\exp[i2\pi \text{CS}_1(\mathcal{A})]$, where
\be
\text{CS}_1(\mathcal{A}) = \frac{i}{2\pi}\int \Tr \mathcal{A}.
\ee
 This is but one  of many topological invariants appearing in the classification of topological phases. See \cite{chiu2015classification} for a detailed and highly accessible  review.

Having  seen an isolated example let us now turn to the classification of all gapped free fermion phases as specified by their dimensions, symmetries and topological properties.  In \cite{PhysRevB.78.195125} Altland and Zirnbauer  classified disordered fermion systems into ten symmetry classes that correspond to the classical  symmetric spaces of compact type. These same ten classes  apply to the Hamiltonians of free fermions systems (irrespective of whether or not they are gapped). The classification of topological phases for gapped free fermion systems was pioneered by Schnyder, Ryu, Furusaki, and Ludwig in \cite{PhysRevB.78.195125} and reviewed in \cite{schnyder2009classification}. In particular, for a given dimension they showed that topologically non-trivial phases  exist in five of the ten symmetry classes. Of these they established that three are furnished with   a $\Z$-valued  invariant, while the remaining two admit a $\Z_2$-valued invariant. Pulling these observations together, Kitaev developed a compelling and influential picture of this classification system, now known as the   ``Periodic Table for topological insulators and superconductors'' \cite{Kitaev:2009mg}. He introduced a set  mathematical organising principles that revealed the Bott periodic features of the classification. In particular, he uncovered the  important role played by Clifford algebras and $K$-theory in understanding the classification of the topological phases.  These ideas have since been developed in number of directions. A systematic and elucidating account of Kitaev's perspective was presented in  \cite{Chiu:2011} and the role of Clifford algebras was emphasised and completed in \cite{abramovici2012clifford}. 
A unified framework on a more  mathematically firm footing was developed by  Freed and Moore in a clear and  comprehensive work \cite{Freed:2012uu}, which, in particular, treated the symmetries carefully and introduced twisted $K$-theory \cite{Atiyah:2004jv} as key tool. Many of the mathematical ambiguities (as well as certain physical puzzles) where clarified in  \cite{Thiang:2014fxa}, again in more rigorous mathematical terms that rely primarily on \emph{unreduced} $K$-theory. We will comment further on some of these and other perspectives once we have introduced more fully  the key ideas put forward in original classification schemes following \cite{PhysRevB.78.195125, Kitaev:2009mg, Chiu:2011}.

Our starting point is the ``ten-fold way'' of Altland and Zirnbauer \cite{altland1997nonstandard}, generalising Dyson's three symmetry classes for many-body systems \cite{dyson1962statistical}.   Consider a generic  matrix $H$,
indices going over position, spin, flavour\ldots, that can be used to build a arbitrary quadratic fermionic  Hamiltonian. The possible matrices $H$ are separated into ten classes under   the presence or absence of   three discrete   transformations $\mathcal{C}$, $\mathcal{T}$ and $\mathcal{P}=\mathcal{C}\cdot\mathcal{T}$ \cite{PhysRevB.78.195125},
\be
\begin{array}{ccc}
\mathcal{C}H\mathcal{C}^{-1}=-H&& \mathcal{C}^2=\pm\mathds{1},\\[5pt]
\mathcal{T}H\mathcal{T}^{-1}=\phantom{-}H&& \mathcal{T}^2=\pm\mathds{1}.
\end{array}
\ee
The particle-hole symmetry (PHS) $\mathcal{C}$ and time-reversal symmetry (TRS) $\mathcal{T}$ are antiunitary operations, while  sublattice symmetry  (SLS) $\mathcal{P}$ is unitary.  In a given basis their action on $H$ can be represented by complex matrices satisfying:
\be
\begin{array}{llllllll}
	C H^*C^{-1} &=& - H	& &C^*C=\pm \mathds{1},\\[5pt]
T H^*T^{-1} &=& \phantom{-}H&	& T^*T=\pm \mathds{1},\\[5pt]
	PHP^{-1} &=& - H&	& P=P^\dagger, P^2= \mathds{1}.	
\end{array}
\ee
For example, recall the BdG Hamiltonian describing superconducting order \eqref{BdG}, 
\be
\mathcal{H}=\Psi^\dagger H \Psi = \frac{1}{2} \begin{pmatrix}\psi^\dagger & \psi \end{pmatrix} \begin{pmatrix}h & \Delta \\ \Delta^\dagger &-h^T\end{pmatrix}\begin{pmatrix}\psi \\ \psi^\dagger \end{pmatrix},\qquad h^\dagger=h, \quad \Delta^T=-\Delta.
\ee
By construction the Nambu spinor $\Psi$  satisfies the constraint
$
\Psi^\dagger = (C\Psi)^T
$
where $C=\sigma_x$ squares to $+\mathds{1}$ and 
\be
CH^*C^{-1}=-H
\ee
so that by definition  the BdG system respects the PHS with $C^*C=+\mathds{1}$,   in this case better referred  to   as the particle-hole \emph{constraint}. Recall, we can rewrite the BdG Hamiltonian in a Majorana basis,
\be
\mathcal{H}=\Gamma^T (iX) \Gamma,  \qquad X^*=X, \quad X^T=-X,
\ee
so that $X$ is an element of $\mathfrak{so}(4n)$.

Examining the various possibilities leads to  ten symmetry classes as shown in \autoref{class}.  There are three possibilities for PHS: no symmetry, symmetry with $\mathcal{C}^2=\mathds{1}$,  and symmetry with $\mathcal{C}^2=-\mathds{1}$. Similarly,  there are three possibilities for TRS, giving a total of nine. Only for the case of no symmetry under both PHS and TRS is the SLS action not fixed. Hence there are ten classes in total.  These are in one-to-one correspondence with the $10=2+8$ classical families of  symmetric spaces of compact type as  classified by Cartan, see \autoref{class}. Again this is easily understood. Roughly speaking the discrete symmetries can be used to construct an  involution, $\theta: \mathfrak{g}\rightarrow\mathfrak{g}, \theta^2=\mathds{1}$, so that,
\be
\mathfrak{g}=\mathfrak{h}+\mathfrak{p}, \quad [\mathfrak{h},\mathfrak{h}]\subset\mathfrak{h},  \quad[\mathfrak{h},\mathfrak{p}]\subset\mathfrak{p}, \quad [\mathfrak{p},\mathfrak{p}]\subset\mathfrak{h}
\ee
where $\mathfrak{h}$  ($\mathfrak{p}$) is the positive (negative) eigenspace of $\theta$. For each symmetry class, specified by the presence/absence of a  PHS/TRS/SLS,  there is an associated  Lie algebra $\mathfrak{g}$ and  involution $\theta$ such that  $iH$ will lie in the corresponding negative eigenspace $\mathfrak{p}$. For example, in the case of no symmetries $H$ is a $n\times n$ Hermitian matrix so that $iH\in \mathfrak{u}(1)\oplus\mathfrak{su}(n)$ and $\exp[iH]$ can be regarded as an element of the type II symmetric space $\Un(n)\times\Un(n)/\Un(n)\cong\Un(n)$ of class $A$ (up to Abelian factors).   The adjoint action of the  SLS operation   squares to identity so that it defines an involution. If   $H$ respects the SLS, that is   anti-commutes with $\mathcal{P}$,  then $iH$ must lie  in the negative eigenspace  $\mathfrak{p}\cong \C^k\otimes\C^m$, where $\mathfrak{u}(n)=\mathfrak{u}(k)\oplus\mathfrak{u}(m)+\mathfrak{p}$ and $n=k+m$. This condition is preserved by the adjoint action of  $\Un(k)\times\Un(m)$ so that $\exp[iH]$ is  an element of the type I symmetric space of class $AIII$ $\Un(n)/\Un(k)\times\Un(m)$. This pair of symmetric spaces, or equivalently symmetry classes, repeats with the order two Bott periodicity of complexified Clifford algebras \eqref{CCliff}.

\newcolumntype{M}{>{$}c<{$}}
\begin{table}[h]
\centering
\begin{tabular}{MM|M|M|M|M|MMMMMMMMMMMMMMMMMMMMMMMMMMMMMMMMM}
 \hline
 \hline
&&&&&&&&&& \\
\text{Class}& q & \text{Coset space} & \text{PHS}& \text{TRS}&  \text{SLS}& 0&1&2&3&4&5&6&7 \\[5pt]
\hline
\hline
&&&&&&&&&&& \\
A&0& \Un(n)\times\Un(n)/\Un(n) &0&0&0& \Z&0&\Z&0&\Z&0&\Z&0\\[5pt]
AIII&1& \Un(n)/\Un(k)\times\Un(m)  &0&0&+&0&\Z&0&\Z&0&\Z&0&\Z\\
&&&&&&&&&&&&& \\
 \hline
 \hline
&&&&&&&&&&&&& \\
D&1& \Orth(n)\times\Orth(n)/\Orth(n)  &+&0&0&\Z_2	&\Z_2	&\Z		&0		&0		&0		&\Z		&0\\[5pt]
DIII&2& \Orth(2n)/\Un(n)    &+&-&+&0		&\Z_2	&\Z_2	&\Z		&0		&0		&0		&\Z\\[5pt]
AII&3& \Un(2n)/\Sp(n) &0&-&0&\Z		&0		&\Z_2	&\Z_2	&\Z		&0		&0		&0\\[5pt]
CII&4& \Sp(n)/\Sp(k)\times\Sp(m) &-&-&+&0		&\Z		&0		&\Z_2	&\Z_2	&\Z		&0		&0\\[5pt]
C&5& \Sp(n)\times\Sp(n)/\Sp(n)  &-&0&0&0		&0		&\Z		&0		&\Z_2	&\Z_2	&\Z		&0\\[5pt]
CI&6& \Sp(n)/\Un(n)  &-&+&+&0		&0		&0		&\Z		&0		&\Z_2	&\Z_2	&\Z\\[5pt]
AI&7& \Un(n)/\Orth(n)  &0&+&0&\Z		&0		&0		&0		&\Z		&0		&\Z_2	&\Z_2\\[5pt]
BDI&0& \Orth(n)/\Orth(k)\times\Orth(m) &+&+&+&\Z_2	&\Z		&0		&0		&0		&\Z		&0		&\Z_2\\[5pt]
&&&&&&&&& \\
 \hline
 \hline
\end{tabular}
\caption{Classification table for topological insulators and superconductors \cite{PhysRevB.78.195125, Kitaev:2009mg, schnyder2009classification}. The first column list the Cartan labels of the symmetric spaces corresponding to the Hamiltonian symmetry classes as determined by the PHS, TRS and SLS columns (here 0 denotes no symmetry and $\pm$ denotes a symmetry that squares to $\pm\mathds{1}$). The second column lists the corresponding  coset spaces, where $n=k+m$. Note, on taking the inductive limit $n\rightarrow\infty$ we obtain the classifying spaces $R_q$ given in \eqref{clc} and \eqref{clr}. The remaining columns contain the classification of  phases  for topological insulators and superconductors with translation invariance in free fermion systems in spatial dimensions 0 through 7. Here a 0 means there is no non-trivial topological phase, while  $\Z_2$ and $\Z$ denote the sets of distinct phases. The real cases are given by the reduced real $K$-theory groups described in \autoref{cliff}, $\widetilde{KO}(S^{q+1-d})\cong \pi_0(R_{q+1-d})$. Note, we have the same Bott periodic pattern in both symmetry class and spatial dimension. We see that the Majorana chain belonging to the $D$ class has   a $\Z_2$ set of phases, one trivial and the other admitting Mzm on the boundaries.}\label{class}
\end{table}

The correspondence between the remaining symmetric spaces and  symmetry classes of $H$ can be made systematic \cite{Chiu:2011} such that they appear in particular order that brings out their mod eight Bott periodic features  first noted in \cite{Kitaev:2009mg}. This approach turns around that of Altland-Zirnbauer by starting with the symmetric spaces, extracting the corresponding Hamiltonian families and only then identifying their symmetry class. Consider $\Orth(16r)$ in the inductive $r\rightarrow \infty$ limit and its vector representation $V\cong\R^{16r}$. Now introduce a set of orthogonal anti-commuting complex structures $J_i$ acting on $V$,
\be
J_iJ_j+J_jJ_i=-2\delta_{ij}\mathds{1},\qquad i,j=1,2,\ldots m,
\ee
constituting a reducible representation of $\Cliff(m,0)$. The subgroup in $\Orth(16r)$ commuting with $J_1$, being a complex structure, is $\Un(8n)$.  Now $J_2$ anti-commutes with the complex structure $J_1$ and so is anti-linear on $\C^{8r}$ and thus defines a quaternionic structure. The subgroup in $\Un(8n)$ commuting with $J_2$ is therefore $\Sp(4r)\cong \Un(4r, \Q)$. Introducing one-by-one further $J_i$ generates a sequence of subgroups repeating modulo eight \cite{Chiu:2011}:
\be\label{groupseq}
\footnotesize
\begin{array}{cccccccccccccccccccc}
\R&&\C&&\Q&&\Q\Q&&\Q&&\C&&\R&&\R\R\\
\Orth(16r)&\subset& \Un(8r) &\subset& \Sp(4r)&\subset&\Sp(2r)\times \Sp(2r) &\subset& \Sp(2r) &\subset& \Un(2r) &\subset& \Orth(2r) &\subset& \Orth(r)\times \Orth(r)
\end{array}
\ee
where we have indicated their corresponding real, complex and quaternionic nature,
\be
\begin{array}{ccccccc}
\mathfrak{so}(n) &\cong& \{x \in \R[n]\,|\, x^\dagger =-x\},\\[5pt]
\mathfrak{u}(n) &\cong& \{x \in \C[n]\,|\, x^\dagger =-x\},\\[5pt]
\mathfrak{sp}(n) &\cong& \{x \in \Q[n]\,|\, x^\dagger =-x\},
\end{array}
\ee
 which reflects the Bott periodic mnemonic \eqref{mn} for the Clifford matrix  algebras.  Let $G$ be a given group in this sequence and $K$ its adjacent subgroup, then $G/K$ is a homogeneous symmetric space. If $G_k$ is the subgroup defined by commuting with all $J_i$ up to $i=k$ then the adjoint action of $J_{k+1}$ on its Lie algebra $\mathfrak{g}_k$ is an involution and $\mathfrak{g}_{k+1}$ (the Lie algebra of $G_{k+1}\equiv K_k$) is its positive eigenspace.  Note, for fixed $J_1,\ldots J_k$ the symmetric space $G_k/G_{k+1}$  parametrises the possible choices of $J_{k+1}$,  since,  essentially, the stability group of $J_{k+1}$ as an element of the adjoint representation of $G_k$ is $G_{k+1}$.  This process gives us an ordered  list of  classical symmetric spaces repeating with order eight,  labelled by $q \in \F_8$ in   \autoref{class}. We note that the exceptional symmetric spaces, which are naturally associated with octonionic geometries \cite{Baez:2001dm}, are not naively applicable since being of fixed size they do not admit a thermodynamic limit. It is perhaps conceivable, although rather unlikely, that a large many-body system could have repeating subsystems constrained by exceptional symmetries, a trace of which might survive the thermodynamic limit. 
 
Given an element in  $\mathfrak{p}\cong T(G/K)$ for a specific $G/K$ we can extract a Hamiltonian and determine its  symmetries \cite{Chiu:2011}.  We have already seen that for the PHS respecting BdG Hamiltonian with $\mathcal{C}^2=\mathds{1}$ and no further symmetries that $iX=H$, where $X\in\mathfrak{so}(4n)$ corresponding to the type II symmetric space in class $D$, $\Orth(4n)\times\Orth(4n)/\Orth(4n)\cong \Orth(4n)$. Conversely, consider an arbitrary element $X\in\mathfrak{so}(4n)$ and map $i\mapsto -i\sigma_y\otimes\mathds{1}$ so that we can identify a Hamiltonian  $H= -i\sigma_y\otimes X$. Then   $\mathcal{C}=\sigma_z \otimes \mathds{1}$ defines a real structure anti-commuting with $H$ and hence a PHS with $\mathcal{C}^2=\mathds{1}$, placing $H$ in symmetry class $D$. Lets move one step along the symmetric spaces to $\Orth(4n)/\Un(2n)$. By construction the elements of $\mathfrak{p}\cong T(\Orth(4n)/\Un(2n))\cong \wedge^2\C^{2n}$ are real anti-symmetric matrices $X\in\wedge^2\R^{4n}$ that anti-commute with a single complex structure $JX=-XJ$. As before let $H= -i\sigma_y\otimes X$ so that $\mathcal{C}=\sigma_z \otimes \mathds{1}$ is again a PHS. Now $\mathcal{C}=\sigma_z\otimes J$ is anti-linear (anti-commutes with $-i\sigma_y\otimes\mathds{1}$), commutes  with $H$ and hence defines a TRS such that $\mathcal{T}^2=-\mathds{1}$. The product $\mathcal{C}\cdot\mathcal{T}$ is given by $\mathds{1}\otimes J$, which is linear (commutes with $-i\sigma_y\otimes\mathds{1}$), anti-commutes with $H$ and  hence defines an SLS. We conclude that $H$ indeed respects the PHS, TRS and SLS  specified by \autoref{class} for class $DIII$. One can continue in this manner for all eight classes \cite{Chiu:2011}. This construction kept things symmetric so that one can continue indefinitely for large enough $r$ and so expose the periodic features, but in general for the relevant cases we can have $G(n)/K(k)\times K(m)$ with $n=k+m$. Note, with the ordering dictated by \eqref{groupseq} the presence of SLS symmetry is alternating while the PHS  follows the same Bott periodic pattern as the symmetric spaces,
 \be
 \begin{array}{cccccccccccccccccccccccccccccccccc}
\R\R&&\R&&\C&&\Q&&\Q\Q&&\Q&&\C&&\R&&\R\R&&\R&&\C&&\Q&&\Q\Q&&\Q\cdots\\
+&& + && 0&&- &&- && - && 0 &&+&&+&& + && 0&&- &&- && -\cdots
\end{array}
 \ee
as does TRS, but shifted back two places. Here we are denoting the absence of the PHS symmetry by 0 and its presence by $\pm$ for $\mathcal{C}^2=\pm\mathds{1}$.

 For the classification of topological phases we are then interested in the partition of a given space $\mathcal{M}$ of admissible $H$ determined  by the symmetry class and  gap condition, which is presumed to form a topological space, into its path-connected components $\pi_0(\mathcal{M})$. Heuristically this corresponds to the equivalence of two Hamiltonians if they can be identified by adiabatic perturbations that do not  close the gap or violate any of the symmetries at any point. If we consider families of Hamiltonians $H(x)$ parametrised by some topological space $x\in X$ (for example a Brillouin zone) then a vector bundle picture merges, typically by associating an isolated family of $m$ negative energy bands with a rank $m$ complex vector bundle, see for example \cite{de2014classification}.  We are then interested in  homotopy equivalent   vector bundles over $X$, as given by homotopy classes of maps from $X$ to some classifying space\footnote{We could have alternatively considered isomorphism classes, then the $K$-theory groups $K(X)$ would arise naturally, but homotopy seems physically  natural and appears to be the correct choice especially when the vector bundles are furnished with extra structure coming from the symmetries classes \cite{de2015chiral}.}. Of course, we have to include the action  of the transformations for a given symmetry class, on both the base space and the fibres, furnishing these vector bundles with additional structure. For example, the TRS of class $AI$ implies the use of  ``real'' vector bundles developed in  \cite{atiyah1966k}. In a series of papers \cite{de2014classification, de2015classification,de2015chiral} the appropriate categories of vector bundles with extra structure where constructed for classes $AI, AII, AIII$ (referred to as  `real', `quaternionic', and `chiral', respectively), and the classification of topological phases for a broad class of compact base spaces was completed. All ten classes where treated in \cite{Kennedy:2015,Kennedy:2014cia} over either spheres or tori equipped with an involution and fibres consisting of complex vector spaces  subject to appropriate constraints reflecting the symmetry class.   Here too we see the Bott periodic features discovery by Kitaev in his original  classification scheme  \cite{Kitaev:2009mg}, which we now discuss.

 Seeking a manageable classification that captures the essential physics Kitaev appealed to the notion of \emph{stable equivalence}, which gives a courser partitioning than homotopy alone and naturally leads us to a $K$-theoretic perspective. The physical picture is that we are allowed to augment our system of interest with additional  degrees of freedom that are typically trivial with respect to the relevant physics and then look for equivalence in this larger space. For example, we may include additional flat bands for an insulator by augmenting with inner atomic shells. Two initially inequivalent phases may become equivalent once we allow for argumentation, while preserving the features we would like to capture in the classification. Kitaev also adopted  (in general) a \emph{relative} perspective, again naturally in keeping with his $K$-theoretic approach. Here one phase is considered relative to another; in essence we classify topological obstructions between Hamiltonians as opposed to the Hamiltonians themselves. The relative approach was also recently advocated in \cite{Thiang:2014wya}, motivated in part by its naturalness, but also because the ``absolute'' picture requires a canonical notion of the trivial or zero phase, which is not always available and can thus lead to certain ambiguities. Indeed, for the  class $AIII$ the equivalence under isomorphism and homotopy in the category of chiral vector bundles do not coincide and the homotopy classification  depends on a choice of reference isomorphism so that only  phases relative to each other  have an absolute meaning \cite{de2015chiral}. 
 
 With these considerations in mind Kitaev defined  two admissible matrices $A,B\in\mathcal{M}$ to be equivalent $A\sim B$ if and only if $A\oplus C$ and   $B\oplus C$ belong to the same element of $\pi_0(\mathcal{M})$ for some matrix $C$ of arbitrary size. Note, $\mathcal{M}$ here places no restriction on the size of admissible matrices. Second, for  a pair of same size  matrices $(A,B)$ he defined the difference class $d(A,B)$   as the equivalence class with representative $(A,B)$ of same size pairs, where two pairs $(A,B)$ and $(A',B')$  are equivalent if and only if   $A\oplus B'\sim A'\oplus B$. Note, this definitions allows for $A$ and $A'$ (and  so $B$ and $B'$) to have different sizes. These definitions capture the notion of augmentation and relativity, respectively. Because we allow for arbitrary size augmentations we might anticipate that the classification for families of Hamiltonians will depend on $[X, R]$, the set of homotopy classes of maps from the parameter space $X$ to  some classifying  space $R$. The action of the symmetries makes this rather subtle, as we shall see. But first let us get a handle on the set of classifying spaces that we expect to be involved.

We have already seen that the symmetry requirements imply that the admissible $A$ (from here on in we will consider $iH$ as opposed to $H$) belong to $\mathfrak{p}\cong T(G/K)$ for $G/K$ belonging to one of the ten classes of symmetric spaces in \autoref{class}. Since we are considering gapped systems Kitaev also required that the admissable $A$ be full 
rank. As we are interested in topological phases the magnitudes of the  eigenvalues are irrelevant;  we can always  use a homotopy equivalent  spectrally flattened matrix satisying $A^2=-\mathds{1}$ with eigenvalues $\pm i$. Let $\mathfrak{p}_l=T(G_l/G_{l+1})$, then for all $A\in \mathfrak{p}_l, A^2=-\mathds{1}$ the matrix $J_{l+1}A$ anticommutes with $J_i, i=1\dots l+1$ and is therefore a candidate $J_{l+2}$. Hence, the set of possible $A\in \mathfrak{p}_l, A^2=-\mathds{1}$ is given by the symmetric space one step along the Bott periodic sequence $G_{l+1}/G_{l+2}$. The space of admissible $A\in\mathfrak{p}_k=T(G_l/G_{l+1})$  of a fixed size is $G_{l+1}/G_{l+2}$, where for the cases $G(n)/K(k)\times K(m)$ we have to take the union of the possible spits $n=k+m$. Hence, allowing for augmentation we expect the sequence   of classifying spaces given by the inductive limits of the $G/K$ given \autoref{class}. For the real order eight case we have $R_q, q\in \F_8$ where,
 \be\label{clr}
 \begin{array}{ccccccccccccccccc}
 R_1 && R_2 && R_3 && R_4 && R_5 && R_6 && R_7 && \Z\times R_0\\[5pt]
 \Orth && \Orth/\Un && \Un/\Sp && \Z \times \text{BSp} && \Sp && \Sp/\Un && \Un/\Orth&& \text{BO}
 \end{array}
 \ee
 For the complex order two case we have $C_q, q\in \F_2$ where 
  \be\label{clc}
 \begin{array}{ccccccccccccccccc}
 C_1 && C_0\\[5pt]
 \Un && \Z\times\text{BU}
 \end{array}
 \ee
For example,  if $X$ is a $d$-sphere with no further structure and we label the symmetry classes (symmetric spaces) by the same $q$, as in \autoref{class},  then for class $q$ we would have $[S^d, R_{q+1}]\cong \pi_d(R_{q+1})\cong \pi_0(q+d+1)$ \cite{Chiu:2011}. In general, however, $X$ will carry extra structure such as involution. 
 
 Kitaev approached the  problem via Clifford algebras and $K$-theory \cite{Kitaev:2009mg}.  As we have seen each set of symmetries corresponding to a given class can be  mapped to a Clifford algebra  $\Cliff(p,q)$ for $q=0$. Moreover, the spectrally flattened $A^2=-\mathds{1}$ can be used to extend the Clifford algebra to $\Cliff(p+1, q)$ and we need to classify the possible extensions in the inductive limit. The key object is the $K$-theory group $K^{p,q}(X)$. Loosely speaking this is the Grothendieck group of  the restriction  functor, induced by the inclusion $\Cliff(p,q)\subset\Cliff(p,q+1)$, from the category of $\Cliff(p,q+1)$-modules to the category of $\Cliff(p,q)$-modules over a base space $X$, which we will assume compact. See \cite{karoubi2008k} chapter II section 2.13 and chapter III section 4. In particular, for   compact $X$ we have the isomorphisms,
 \be
 K_{\R}^{p,q}(X)\cong [X, R_{q-p}],
 \ee
 where $q-p$ is taken module 8, and 
  \be
 K_{\C}^{p,q}(X)\cong [X, C_{q-p}],
 \ee
where $q-p$ is taken module 2 \cite{karoubi2008k}. Note, for the problem at hand we can (and Kitaev did)   use the Clifford period relations,
\be
\begin{split}
\Cliff(0,d+2)&\cong \Cliff(d,0)\otimes \Cliff(0,2)\\
\Cliff(d+2,0)&\cong \Cliff(0,d)\otimes \Cliff(2,0)\\
\Cliff(s,t)&\cong \Cliff(s,t)\otimes \Cliff(1,1)\\
\end{split}
\ee
to send  $p\rightarrow 0$ and $q\rightarrow p+2$ up to Morita equivalence.

For translationally invariant systems in an arbitrary spatial dimension $d$ we can consider the family of $A(k)$ for $k\in\tilde{\R}^d$, where the tilde indicates standard Euclidean space equipped with an involution $\tau(k)=-k$. The $K$-theory of  `real' vector bundles over a topological space $X$ with an involution $\tau:X\rightarrow X, \tau^2=\mathds{1}$ was developed by Atiyah \cite{atiyah1966k}, where it was denoted $KR(X)$. See also \cite{de2014classification} for a rigorous treatment of `real' vector bundles applied to class $AI$. To obtain a reasonable classification Kitaev considered fixed  $|k|\rightarrow\infty$ asymptotics for $A(k)$ so that we can  regard infinity   as the boundary $\tilde{S}^{d-1}$ of a large ball $\tilde{B}^d$. Then for real symmetry class $q$ in $d$ spatial dimensions the classification is given by the relative\footnote{ A relative $K$-theory group $K(X, Y)$, for $Y$ closed in $X$,  is  the Grothendieck group of the functor induced by the restriction of bundles on $X$ to those on  $Y$.  Note, we have an isomorphism with the standard reduced $K$-theory, $K(X, Y)\cong \widetilde{K}(X/Y)$, which is sometimes taken as the definition.  $K^{p,q}(X, Y)$ is the appropriate corresponding generalisation of $K^{p,q}(X)$.  See \cite{karoubi2008k} chapter II section 2 and chapter III section 5.} $K$-theory group \cite{Kitaev:2009mg},
\be
K_{\R}^{0,q+1}(\tilde{B}^d, \tilde{S}^{d-1})\cong \pi_0(R_{q+1-d})\cong \widetilde{KO}(S^{q+1-d}),
\ee
which yields \autoref{class} as in \cite{Kitaev:2009mg, schnyder2009classification}. The $d=1$ Majorana chain from \autoref{mc} belongs to class $D$ $(q=1)$ and is thus predicted correctly to have two phases $\pi_0(R_{1})\cong\Z_2$, one of which is topologically non-trivial, or alternatively a single $\Z_2$-valued invariant separating the phases. To give another example,  topological superconductors in $d=1$ and  class $DIII$ (as considered, for example, in the context of B phase  superfluid ${}^3$He \cite{PhysRevB.78.195125}) have phases labeled by a $\Z$-valued topological invariant,   the winding number. For a comprehensive review of \autoref{class} along with various examples realising its entries see \cite{chiu2015classification}. We should mention that there are a number of subtleties hidden by \autoref{class}. For example, the use of stable equivalence has led to the prediction of no topological phases in $d=3$ for class $A$, yet if one restricts to two bands  then non-trivial phases appear \cite{Kennedy:2014cia}, although they are unstable under the addition of trivial bands.  Finally, let us conclude by noting that the pattern appearing  in \autoref{class} can be reinterpreted in terms of the division algebras by passing through Clifford algebras as in \autoref{cliff}. On the surface at least this appears to be nothing more than curiosity; whether or not the division algebras could play a more direct role is left for future consideration.   

\section{Conclusions}

We have touched on  Majorana's impact in particle physics, condensed matter and quantum computation. In \autoref{intro} we reviewed   Majorana's  original vision, which may yet prove crucial to our understanding of  neutrino physics.  We then moved on to the role of Majorana quasiparticles as emergent  states appearing in condensed matter systems with superconducting order in \autoref{MCM}. This included both Majorana fermions and their more exotic cousins,  Majorana zero-modes  exhibiting non-Abelian ayononic statistics. We indicated how attributes  of the latter lend themselves to fault-tolerant topological quantum computation. To close we examined the classification of gapped free fermion phases in condensed matter, which tied together our understanding of many-body systems and the theory of spacetime spinors appearing  in fundamental particle physics via Clifford algebras and $K$-theory.  

The impact of topological phases in quantum computing represents a marvellous and rather rare example of fundamental theoretical advances progressing hand-in-hand with their technological applications \cite{Nayak:2008zza}. One should also keep in mind, however, that this synthesis  has its germ in  numerous fundamental results of  curiosity driven science, stretching from  Majorana to today's cutting edge research in condensed matter. The pioneers of these varied  fields could have scarcely   anticipated their current technological applications, let alone have been motivated by them. This brings us back to the sentiment expressed in the closing remarks of \autoref{intro}.  Good science seeking to expand our basic understanding needs no other  motivation;  rewards of a broader nature surely  will follow. The intrinsic quality of this form of technological progress is in the very fact that it cannot be anticipated in the present.

Our survey of Majorana fermions has a glaring omission: the crucial role they play in supersymmetry, supergravity, string theory and M-theory. This was left to \cite{Ferrara:2015bqa}. However,  we cannot resist concluding with some remarks that connect to  aspects of  string/M-theory. A crucial feature of topological phases that we have not properly touched on, although Mzm in the open Kitaev chain are an explicit example, is the bulk-boundary correspondence. See \cite{PhysRevB.78.195125,chiu2015classification} and the references therein. Topologically trivial and non-trivial phases must be separated by a quantum phases transition. The implication is that if we put together two  $d$-dimensional systems, one in a trivial phase and the other in a non-trivial phase, then the $(d-1)$-dimensional interface must support gapless states, typically Dirac or Majorana zero-modes. These states are topologically protected against perturbations as long as the bulk gap and symmetries are preserved.

This bulk-boundary picture leads us to another perspective: topological phases can be characterised by anomalies \cite{Chen:2011pg, Witten:2015aba}.  A  theory well-defined on a $d$-dimensional spatial manifold without boundary, is not necessarily well-defined when a boundary is introduced. To construct a sensible theory  some additional degrees  of freedom must be introduced on the boundary. Particularly relevant here is the case when such degrees of freedom are gapless. The boundary theory must have an anomaly in order to cancel that of the bulk theory; this is the idea of anomaly inflow \cite{Callan:1984sa}.   The standard example in condensed matter  is the integer quantum Hall effect in two spatial dimensions. The spacetime theory has  a Chern-Simons coupling of the gauge field, which is not gauge invariant when there is a boundary. This deficiency is compensated by the existence of chiral modes on the boundary. These same ideas have been be applied to a significant proportion of the fermionic  topological phases  considered in \autoref{class} \cite{Ludwig:2012}. In particular, going beyond ``global anomalies''  it was shown in  \cite{Witten:2015aba} using the Dia-Freed theorem   that the phases of a 3-dimensional topological superconductor in class $DIII$, where the boundary theory consists of gapless Majorana fermions, are reduced from $\Z$ to $\Z_{16}$, as one would have deduced by introducing   interactions \cite{Fidkowski:2013}. That it is 16 and not 8, as  would have been predicted using the absence of anomalies alone,  is rather subtle.  Although there is no traditional anomaly,  there is an ambiguity in the sign of the  partition function that cannot be consistently resolved  unless the number of fermions is a multiple of 16 \cite{Witten:2015aba}. The root of this issue is that, even when there are no global anomalies, it may not be possible to consistently define the overall phase of the fermion partition function on all  manifolds, generally leading to a violation of unitarity and gluing. 

This  ``anomaly framework'' makes use of a relativistic formulation, which is possible because the boundary fermions have an emergent relativistic symmetry. On the one hand, reliance on a relativistic framework might be regarded as unnatural for condensed matter systems, but on the other it offers an intriguing perspective on genuinely relativistic systems. For example, one could ask for a condensed matter analog of the M2-M5 brane system in M-theory as was done in \cite{Witten:2015}. The partial answer given there was a topological superconductor with its boundary split into two regions, one of which has as usual gapless modes, while the other has symmetry respecting gapped modes.   

The sharing of ideas between condensed matter and string/M-theory is not a one-way street.  A key idea underpinning  M-theory is that of U-duality, a set of quantum dualities that interchange dual descriptions of the same physics. The best understood class of such operations go by the name of T-duality, which for example interchanges type IIA and type IIB string theories.  It turns out that T-duality admits a direct analog for topological insulators \cite{Hannabuss:2016hnw} that simplifies the bulk-boundary correspondence!

\section*{Acknowledgments}

MJD is grateful to the Leverhulme Trust for an Emeritus Fellowship, to Prof Zichichi for his hospitality and to Philip Candelas for hospitality at the Mathematical Institute, Oxford. LB is grateful to Graham Kells  for informative discussions and comments. The work of LB is supported by a Schr\"{o}dinger Fellowship. The work of MJD is supported by the STFC under rolling grant ST/G000743/1.

\appendix

\section{Division algebras}\label{div}
 An algebra $\mathds{A}$ defined over  $\R$ with identity element $e_0$, is said to be \emph{composition} if it has a non-degenerate quadratic form\footnote{A \emph{quadratic norm} on a vector space $V$ over a field $\R$ is a map $\bn:V\to\R$ such that: (1) $\bn(\lambda a)=\lambda^2\bn(a),  \lambda\in\R, a\in V$ and 
(2)
$
\langle a, b\rangle:=\bn(a+b)-\bn(a)-\bn(b)
$
is bilinear.}
$\bn:\mathds{A}\to\R$ such that,
\begin{equation}
\bn(ab)=\bn(a)\bn(b),\quad \forall~~ a,b \in\alg,
\end{equation}
where we denote the multiplicative product of the algebra by juxtaposition.  Regarding ${\R}\subset\alg$ as the scalar multiples of the identity $ \R e_0$ we may decompose $\mathds{A}$ into its ``real'' and ``imaginary'' parts $\alg={\R}\oplus \alg'$, where $\alg'\subset\alg$ is the subspace orthogonal to $\R$. An arbitrary element $a\in\alg$ may be written $a=\text{Re}(a) +\text{Im}(a)$. Here  $\text{Re}(a)\in\R e_0$, $\text{Im}(a)\in \alg'$ and
\be
\text{Re}(a)=\frac{1}{2}(a+\overline{a}), \qquad \text{Im}(a)=\frac{1}{2}(a-\overline{a}),
\ee
where we have defined conjugation  using the bilinear form,
\be
\overline{a}:=\blf{a}{e_0}e_0-a, \qquad \langle a, b\rangle:=\bn(a+b)-\bn(a)-\bn(b).
\ee 

A composition algebra $\mathds{A}$  is said to be \emph{division} if it contains no zero divisors,
\begin{equation*}
ab=0\quad \Rightarrow\quad a=0\quad\text{or}\quad b=0,
\end{equation*}
in which case $\bn$ is positive semi-definite and $\alg$ is referred to as a normed division algebra. Hurwitz's celebrated theorem states that there are exactly four normed division algebras \cite{Hurwitz:1898}: the reals, complexes, quaternions and octonions, denoted respectively by $\R, \C, \Q$ and $\Oct$. They may be constructed via the Cayley-Dickson doubling procedure, $\alg'=\alg\oplus\alg$  with multiplication in $\alg'$ defined by
\be
(a, b)(c, d) = (ac - d\bar{b}, \bar{a}d + cb).
\ee
With each doubling a property is lost as summarised here:
\[
\begin{array}{lllllll}
\alg & Construction & Dim& Division & Associative & Commutative & Ordered \\
\R & \R & 1& yes & yes & yes & yes \\
\C &\R \oplus \R & 2& yes & yes & yes & no \\
\Q &\C \oplus  \C & 4& yes & yes & no & no \\
\Oct &\Q \oplus  \Q & 8& yes & no & no & no \\
\mathds{S} & \Oct\oplus  \Oct & 16 & no & no & no & no \\
\end{array}
\]
On doubling the octonions, $\mathds{S} \cong\Oct\oplus  \Oct$,  the division property fails and we will not consider such cases here. Note that, while the octonions are not associative they are alternative:
\be
[a, b, c]:= (ab)c-a(bc)
\ee
is an alternating function under the interchange of its arguments. This property is crucial for supersymmetry.

\begin{figure}[h!]
  \centering
    \includegraphics[width=0.4\textwidth]{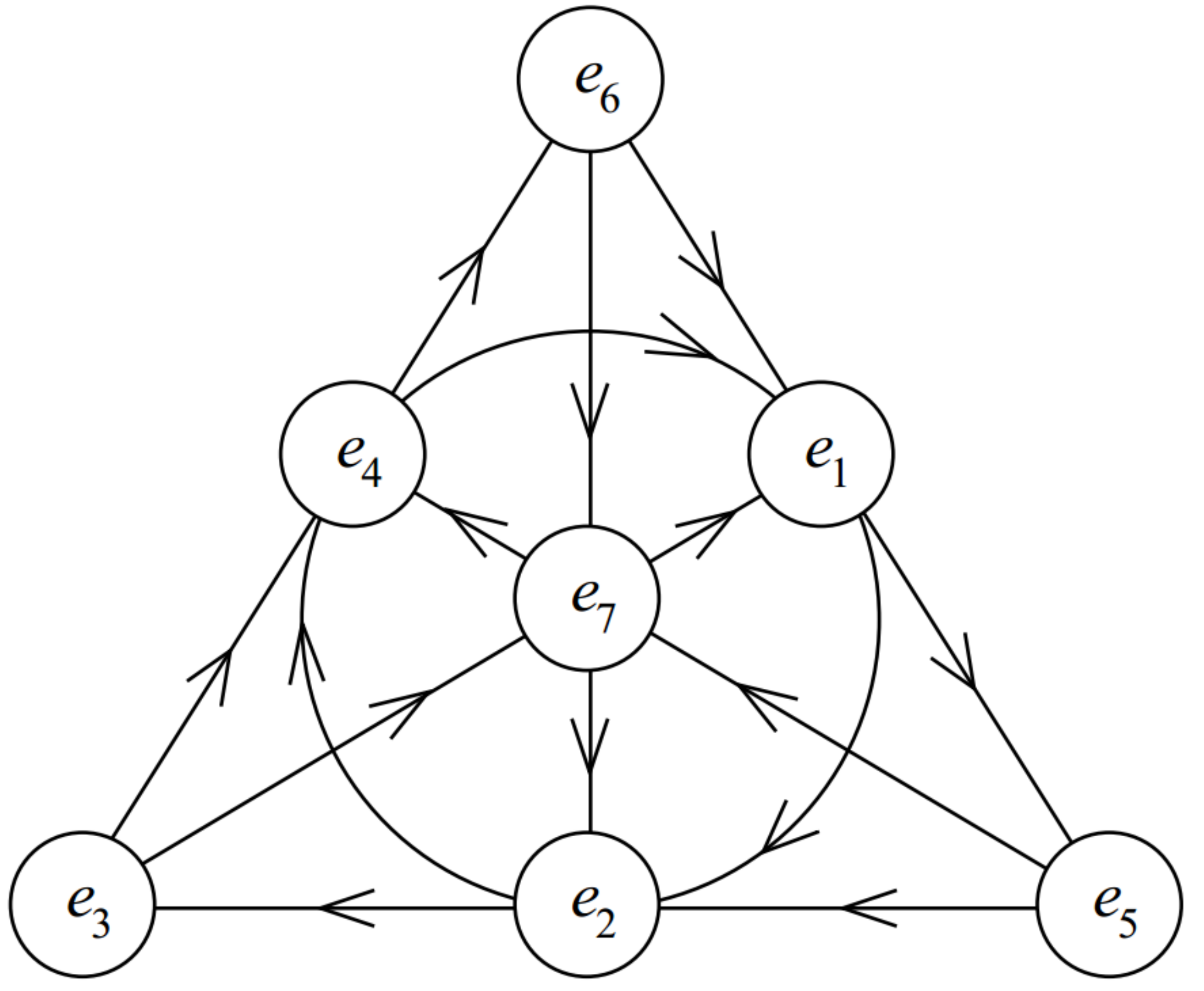}
  \caption{\footnotesize{The Fano plane. The structure constants are determined by the Fano plane, $C_{ijk}=1$ if $ijk$ lies on a line and is ordered according as its orientation. Each oriented line follows the rules of quaternionic multiplication. For example, $e_2e_3=e_5$ and cyclic permutations; odd permutations go against the direction of the arrows on the Fano plane and we pick up a minus sign, e.g. $e_3e_2=-e_5$.}}\label{FANO}
\end{figure}

An element $a\in\Oct$ may be written $a=a^ae_a$, where $a=0,\ldots,7$,  $a^a\in \R$ and $\{e_a\}$ is a basis with one real $e_0$ and  seven $e_i, i=1,\ldots, 7,$ imaginary elements. The octonionic multiplication rule is,
\be
e_ae_b=\left(\delta_{a0}\delta_{bc}+\delta_{0b}\delta_{ac}-\delta_{ab}\delta_{0c}+C_{abc}\right)e_c,
\ee
where $C_{abc}$ is totally antisymmetric and $C_{0bc}=0$.
The non-zero $C_{ijk}$  are given by the Fano plane. See \autoref{FANO}.


\providecommand{\href}[2]{#2}\begingroup\raggedright\endgroup

\end{document}